\documentclass[aps,prd,preprint,amsmath,citeautoscript,longbibliography,nofootinbib]{revtex4-1}
\usepackage[utf8]{inputenc}
\usepackage{bm}
\usepackage{amsmath,mathrsfs,amsfonts}
\usepackage{graphicx,afterpage}
\usepackage{subcaption}
\usepackage{amsmath,latexsym}
\usepackage{color,soul}
\usepackage{float}
\usepackage[section]{placeins}
\usepackage{silence}
\begin{document}
\title{Exact Dynamical Black Hole Solutions in Five or Higher Dimensions}
\author{Bardia H. Fahim{\footnote{bardia.fahim@usask.ca}} and A. M. Ghezelbash{\footnote{masoud.ghezelbash@usask.ca}}}

\affiliation{Department of Physics and Engineering Physics, University of Saskatchewan, Saskatoon SK S7N 5E2, Canada}
\date{\today}
\begin{abstract}
  We construct new classes of the dynamical black hole solutions in five or higher dimensional Einstein-Maxwell theory, coupled to a dilaton field, in the presence of arbitrary cosmological constant. The dilaton field interacts non-trivially with the Maxwell field, as well as the cosmological constant, with two arbitrary coupling constants. The solutions are non-stationary, and almost conformally regular everywhere.  To construct the solutions, we use the four-dimensional Bianchi type IX geometry, as the base space.
We find three different classes of solutions, based on the values of the coupling constants. We notice that our solutions could be asymptotically de-Sitter, anti-de-Sitter or flat. We find the relevant quantities of the solutions, and discuss the properties of the solutions. 
\end{abstract}

\maketitle

\newpage
\section{Introduction}
\label{sec:intro}
Finding the exact solutions to the Einstein gravity, especially in the presence of matter fields in different dimensions, is truly the main aim of gravitational physics. In this regard, the dimensional compactification of higher dimensional gravity has been studied broadly in different references \cite{freund1980dynamics,kaluza2018unification}.
In other area of research on holography between two different models of physics, especially AdS/CFT and Kerr/CFT correspondences, 
constructing the exact solutions to the asymptotically de-Sitter and Anti-de-Sitter Einstein gravity is of utmost importance \cite{guica2009kerr}.  Moreover, the exact solutions to the Einstein gravity coupled to the different matter fields, such as Maxwell field, dilaton field and NUT charges \textcolor{black}{are} constructed in \cite{ghezelbash2015cosmological}-\cite{mahapatra2018time}. 
The Einstein-Maxwell-dilaton theory with different types of interaction between the fields, can describe physical phenomenons, such as slowly rotation black holes \cite{stetsko2019slowly}, {topological charged hairy black holes \cite{add1},} cosmic censorship \cite{goulart2018violation}, gravitational radiation \cite{julie2018gravitational}, hyperscaling violation \cite{li2017hyperscaling} and compactification of M-theory in generalized Freund-Rubin theory \cite{torii2003cosmological}. The Einstein-Maxwell-dilaton theory and its extensions have also been used in other areas of research on black holes \cite{Rec1}-\cite{Rec7}. 

In a recent work \cite{add2}, the authors considered the Einstein-Maxwell-dilaton theory with two extra vector fields. The first field supports the non-trivial topology, and the second field supports states with the finite charge density.  They found a new class of charged black holes with hyper-scaling violating asymptotics. The black holes have non-trivial horizon topology, for arbitrary Lifshitz exponent and a hyper-scaling violation parameter \cite{add2}. 

Moreover, in \cite{NEW15}-\cite{NEW154}, the authors constructed time-dependent charged black hole solutions in different dimensions in Einstein-Maxwell theory. Some of the time-dependent solutions can describe the coalescence of the extremal charged black holes in different dimensions.

After direct detection of  coalescing black holes \cite{BHS}, there is a huge interest in finding the exact analytical dynamical black hole solutions. In general, finding the exact analytical dynamical black hole solutions is a difficult task in general relativity (and its modified versions which include the matter fields). 


Inspired by above considerations, in this article we generalize the Kastor-Traschen-like  black holes \cite{26, 27} in any dimensions $D \geq 5$, based on an embedded four-dimensional Bianchi type IX space. The Kastor-Traschen black holes on Gibbons-Hawking space were constructed in \cite{37, 38}, which describe a system of coalescing black holes.  

The organization of the paper is as follows.  In section \ref{NOTEQ}, we briefly  review the Bianchi type IX space and discuss its properties. Then, we consider the Einstein-Maxwell-dilaton theory in $N+1$-dimensions with two non-equal coupling constants. We employ special ansatzes for the metric, the Maxwell field and the dilaton, and explicitly solve all the field equations. We find analytical exact solutions for all the metric functions, the Maxwell field and the dilaton field. Moreover, we find a constraint on the two non-equal coupling constants and also another constraint on the cosmological constant. We discuss and plot the quantities related to the geometry. In section \ref{sec:equal}, we consider the Einstein-Maxwell-dilaton theory in $N+1$-dimensions with two non-zero equal coupling constants. We employ another different set of ansatzes for the metric, the Maxwell field and the dilaton, and explicitly solve all the field equations. We find analytical exact solutions for all the metric functions, the Maxwell field and the dilaton field. Moreover, we find a constraint on the cosmological constant. We discuss and plot the quantities related to the geometry. In section \ref{sec:00}, we consider the Einstein-Maxwell-dilaton theory in $N+1$-dimensions with two zero coupling constants. The theory reduces to the Einstein-Maxwell theory. We employ special ansatzes for the metric and the Maxwell field, and explicitly solve all the field equations. We find analytical exact solutions for all the metric functions and the Maxwell field.  We discuss and plot the quantities related to the geometry.
In section \ref{sec:up}, we show that the solutions to the Einstein-Maxwell-dilaton theory in section \ref{NOTEQ} can be embedded to a higher-dimensional gravity theory coupled to a form field. We find that the dimension of the the internal space is related to the coupling constant $b$ in the Einstein-Maxwell-dilaton theory.
We wrap up the article by an appendix and the concluding remarks and comments on the future works.
\section{Bianchi type IX geometry and Einstein-Maxwell-dilaton theory with two non-equal coupling constants}
\label{NOTEQ}
In Bianchi's classification of the homogeneous spaces, Bianchi type IX is a self-dual and asymptotically Euclidean space and includes two important sub-spaces such as Eguchi-Hanson I and II. The Bianchi type IX geometry with an $SU(2)$ isometry group is given by
\begin{equation}
ds^2_{}=e^{2f(\eta)}\sigma^2_1+e^{2h(\eta)}\sigma^2_2+e^{2g(\eta)}\sigma^2_3 + e^{2(f(\eta)+h(\eta)+g(\eta))}d\eta^2, \label{BIX}
\end{equation}
where the Maurer-Cartan one-forms $\sigma_i$ satisfy $d\sigma_i=\frac{1}{2}\epsilon_{ijk}\sigma_j\sigma_k$, and are given by
\begin{equation}
    \sigma_1=d\psi+\cos{\theta}d\phi, \label{sig1}
\end{equation}
\begin{equation}
    \sigma_2=\cos{\psi} \sin{\theta}d\phi-\sin{\psi}d\theta, \label{sig2}
\end{equation}
\begin{equation}
    \sigma_3=\sin{\psi}\sin{\theta}d\phi+\cos{\psi}d\theta. \label{sig3}
\end{equation}

The periodicity of the Euler angles $\theta$, $\phi$ and $\psi$ are $\pi$, $2\pi$ and $4\pi$, respectively. From the self-duality property of Bianchi type IX, we find
\begin{equation}
    2\frac{df}{d\eta}=e^{2h}+e^{2g}-e^{2f}-2\lambda_1e^{h+g}, \label{l1}
\end{equation}
\begin{equation}
    2\frac{dh}{d\eta}=e^{2g}+e^{2f}-e^{2h}-2\lambda_2e^{f+g}, \label{l2}
\end{equation}
\begin{equation}
    2\frac{dg}{d\eta}=e^{2f}+e^{2h}-e^{2g}-2\lambda_3e^{f+h}, \label{l3}
\end{equation}
where the constants $\lambda_i$ satisfy $\lambda_i \lambda_j=\epsilon_{ijk}\lambda_k$, with $i=1,2,3$. By choosing $(\lambda_1,\lambda_2,\lambda_3)=(0,0,0)$ and solving the vacuum Einstein equations, we find the functions $f(\eta)$, $h(\eta)$ and $g(\eta)$ in terms of the standard Jacobi elliptic functions $\textbf{sn}$, $\textbf{cn}$ and $\textbf{dn}$ as \cite{ghezelbash2008supergravity,meyer2001jacobi}

\begin{equation}
    f(\eta)=\frac{1}{2}\ln(c^2\frac{\textbf{cn}(c^2\eta ,k^2)\textbf{dn}(c^2\eta ,k^2)}{\textbf{sn}(-c^2\eta ,k^2)}),
\end{equation}
\begin{equation}
    h(\eta)=\frac{1}{2}\ln(c^2\frac{\textbf{cn}(c^2\eta,k^2)}{\textbf{dn}(c^2\eta,k^2)\textbf{sn}(-c^2\eta,k^2)}),
\end{equation}
\begin{equation}
    g(\eta)=\frac{1}{2}\ln(c^2\frac{\textbf{dn}(c^2\eta,k^2)}{\textbf{cn}(c^2\eta,k^2)\textbf{sn}(-c^2\eta,k^2)}).
\end{equation}

Changing the coordinate $\eta$ to $r=\frac{2c}{\sqrt{\textbf{sn}(c^2\eta,k^2)}}$, we find the triaxial Bianchi type IX geometry
\begin{eqnarray}
     ds^2_{B.IX} &=& \frac{dr^2}{\sqrt{g(r)}}+\frac{r^2}{4}\sqrt{g(r)}\{\frac{(d\psi+\cos{\theta}d\phi)^2}{1-\frac{a_1^4}{r^4}} \nonumber\\
     &+& \, \frac{(-\sin{\psi}d\theta+\cos{\psi}\sin{\theta}d\phi)^2}{1-\frac{a_2^4}{r^4}}+\frac{(\cos{\psi}d\theta+\sin{\psi}\sin{\theta}d\phi)^2}{1-\frac{a_3^4}{r^4}}\}, \label{trBIX}
\end{eqnarray}
where 
\begin{equation}
      g(r)=(1-\frac{a_1^4}{r^4})(1-\frac{a_2^4}{r^4})(1-\frac{a_3^4}{r^4}), \label{Jbix}
\end{equation}
and we can choose the parameters $a_1=0$, $a_2=2kc$ and $a_3=2c$, with constants $c>0$ and $0 \leq k \leq 1$. In order to preserve the positive definiteness of the metric (\ref{trBIX}), we need to impose $r\geq a_3$. Bianchi type IX is an important geometry and has been used in different theories such as supergravity, loop quantum cosmology and string theory \cite{bali2001bianchi}-\cite{starobinsky2020anisotropy}.

Having constructed a background space for the theory, we focus on the action of the Einstein-Maxwell-dilaton theory, in $N+1$ dimensions, where the dilaton field is coupled to the electromagnetic field and the cosmological constant, with two different coupling constants $a$ and $b$ \cite{maki1993multi}
\begin{equation}
    S=\int d^{N+1}x\sqrt{-g}\{R-\frac{4}{N-1}(\nabla\phi)^2-e^{-4/(N-1)a\phi}F^2-e^{4/(N-1)b\phi}\Lambda\}, \label{action}
\end{equation}
where $R$ is the Ricci scalar, $F_{\mu\nu}=\nabla_{\mu}A_{\nu}-\nabla_{\nu}A_{\mu}$ is the electromagnetic field tensor, $\phi$ is the dilaton field and $\Lambda$ the cosmological constant. We consider the following ansatz for the $N+1$-dimensional metric
as a background geometry
\begin{equation}
    ds_{N+1}^2=-\frac{1}{H(r,\theta)^2}dt^2+H(r,\theta)^{\frac{2}{(N-2)}}R(t)^2[ds_{B. IX}^2+\Sigma_{i=1}^{N-4}dx^2_i], \label{st}
\end{equation}
where $R(t)$ and $H(r,\theta)$ are two metric functions, $\Sigma_{i=1}^{N-4}dx^2_i$ is the extended Euclidean spaces, and $ds^2_{B.IX}$ is the four-dimensional Bianchi type IX metric given by (\ref{trBIX}).

Variation of the action (\ref{action}) with respect to the metric tensor $g_{\mu\nu}$, electromagnetic gauge field $A_{\mu}$ and the dilaton field $\phi$ leads to the Einstein field equations, electromagnetic and dilaton field equations in $N+1$ dimensions \cite{maki1993multi},
\begin{eqnarray}
    \mathcal{G}_{\mu \nu} &\equiv& R_{\mu \nu}-\frac{1}{2}g_{\mu \nu} R-\frac{4}{N-1}[\nabla_\mu\phi\nabla_\nu\phi-\frac{1}{2}g_{\mu \nu}(\nabla\phi)^2]-e^{\frac{-4a\phi}{N-1}}[2F_{\mu \rho} F_\nu \ ^\rho-\frac{1}{2}g_{\mu \nu} (F)^2]\nonumber\\
    &+& \, \frac{1}{2}e^{\frac{4b\phi}{N-1}}g_{\mu \nu}\Lambda{=0},\label{ein}
\end{eqnarray}
\begin{equation}
    \mathcal{M}_\mu\equiv \nabla^\nu(e^{-4/(N-1)a\phi}F_{\mu \nu})=0, \label{maxwell}
\end{equation}
\begin{equation}
    {\cal D}\equiv \nabla^2\phi-\frac{b}{2}e^{4/(N-1)b\phi}\Lambda +\frac{a}{2}e^{-4/(N-1)a\phi}F^2=0. \label{dilaton}
\end{equation}
respectively. 

We consider the following ansatz for the dilaton field
\begin{equation}
    \phi(t,r,\theta)=-\frac{(N-1)}{4a}\ln{(H^U(r,\theta)R^V(t))}, \label{dil}
\end{equation}
where $U$ and $V$ are two constants that will be determined throughout  solving the field equations of motion. We also consider the following ansatz for the electromagnetic gauge field
\begin{equation}
  A_t(t,r,\theta)=\alpha R^X(t)H^Y(r,\theta), \label{A}
\end{equation}
where we assumed $A_t(t,r,\theta)$ to be the only non-zero component of the electromagnetic gauge field, which generates an electric field in $r$ and $\theta$ directions. In (\ref{A}), $\alpha$, $X$ and $Y$ are constants.

The $\mathcal{M}_r$ component of the electromagnetic field equation (\ref{maxwell}) gives the following differential equation
\begin{eqnarray}
    \mathcal{M}_r &=& -\sqrt{\frac{r^8-16c^4(k^4+1)r^4+256c^8k^4}{r^8}}(\partial_r H)(\partial_t R)Y\alpha (X+V+N-2) \label{mrc}
    \\ \nonumber
    &\times& R^{X+V-3}H^{\frac{N-4}{N-2}+Y+U}=0,
\end{eqnarray}
where $N$ is the number of spatial dimension. This equation (\ref{mrc}) leads to the following relation between the constants $X$ and $V$
\begin{equation}
    X+V=2-N. \label{c1}
\end{equation}

\begin{figure}[H]
    \centering
    \includegraphics[scale=0.40]{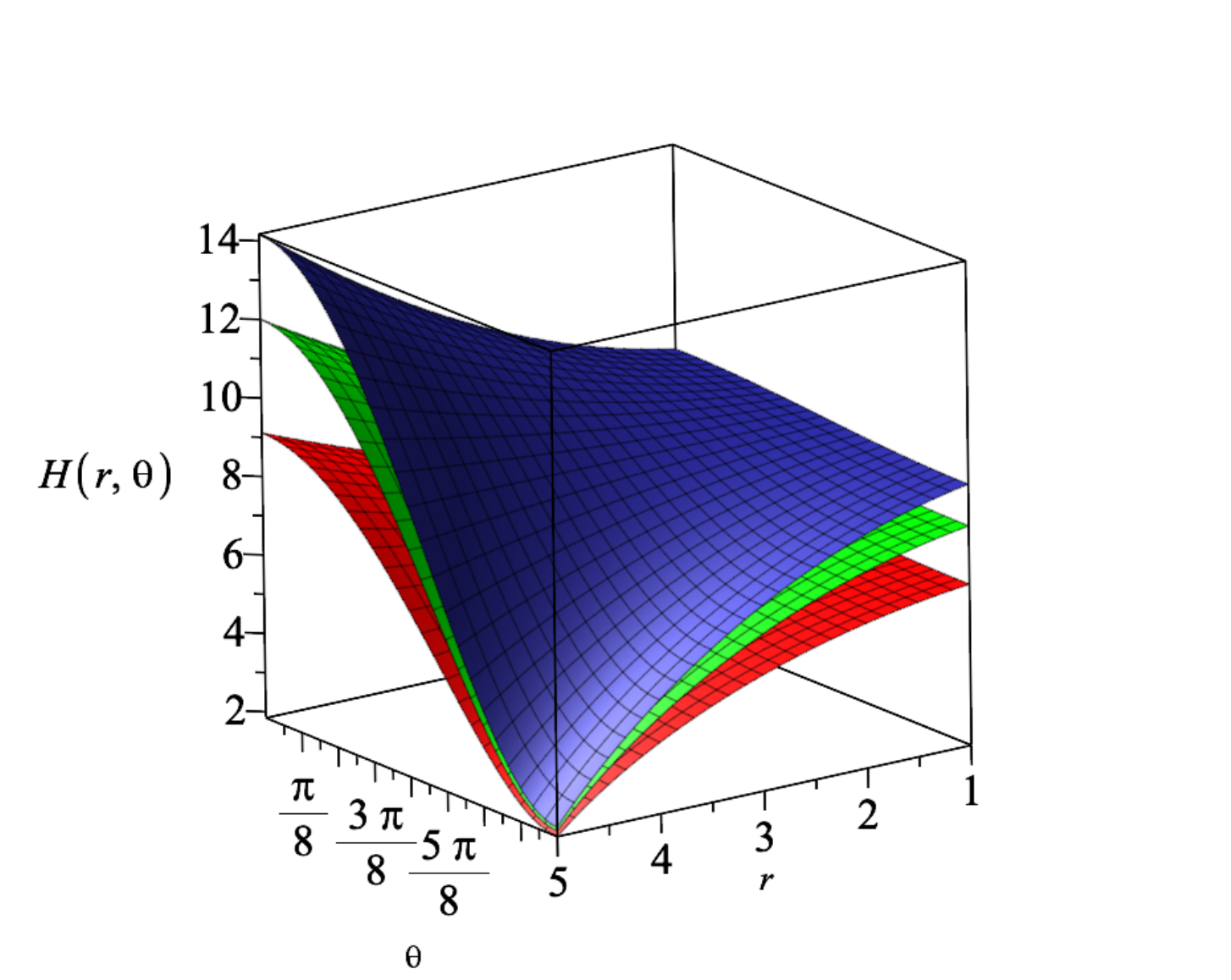}
    \caption{The metric function $H(r,\theta)$ in terms of the coordinates $r$ and $\theta$ for three different spatial dimensions $N=4$, $N=5$ and $N=6$, which correspond to the lower, middle and upper surface, respectively. We assumed the constants $g_+=0.5$, $g_-=15$ and $a=1$. }
    \label{figure:H}
\end{figure}

Moreover, from the $\mathcal{G}_{tr}$ component of the Einstein field equation
\begin{equation}
    \mathcal{G}_{tr}=-\frac{(N-1)\partial_tR \partial_rH(UV+4a^2)}{8a^2HR},
\end{equation}
we find
\begin{equation}
    UV=-4a^2. \label{c2}
\end{equation}
Substituting these constraints (\ref{c1}) and (\ref{c2}) in $\mathcal{G}_{r\theta}$ component of the Einstein field equation, we can determine the constants that appear in dilaton field (\ref{dil})
\begin{equation}
     U=\frac{2a^2}{N-2},  V=-2(N-2),
\end{equation}
and the constants in the gauge ansatz (\ref{A})
\begin{equation}
     X=N-2, Y=-1-\frac{a^2}{N-2},  \alpha^2=\frac{N-1}{2(a^2+N-2)}.
\end{equation}
Substituting these results into the $\mathcal{M}_t$ component of the Maxwell field equation (which due to its length, we show it explicitly in the appendix) gives a differential equation for the metric function $H(r,\theta)$, that can be solved as
\begin{equation}
H(r,\theta)=(g_+r^2\cos{\theta}+g_-)^{\frac{N-2}{a^2+N-2}}, \label{H}
\end{equation}
 where $g_{\pm}$ are arbitrary constants. This suggest that for a fixed dimension, increasing the coupling constant $|a|$, decreases the value of $H(r,\theta)$. We also note that based on the line element ansatz in equation (\ref{st}), the function $H(r,\theta)$ needs to be a real-positive function. In figure \ref{figure:H}, we represent the behaviour of the metric function $H(r,\theta)$ with respect to the coordinates $r$ and $\theta$, for three different dimensions, where we set the coupling constant $a=1$.

 By solving $\mathcal{G}_{rr}$ and $\mathcal{G}_{tt}$ components of Einstein field equations, we find  the following solutions for the metric function $R(t)$ and the cosmological constant $\Lambda$
\begin{equation}
    R(t)=(\eta t+\nu)^{\frac{a^2}{(N-2)^2}},\label{RR}
\end{equation}
\begin{equation}
    \Lambda=\frac{N-1}{(N-2)^2}a^2\eta^2(\frac{Na^2}{(N-2)^2}-1), \label{cosmoc}
\end{equation}
where $\eta$ and $\nu$ are arbitrary constants, $N$ is the number of spatial dimensions and $a$ is the coupling constant. We can avoid $R(t)=0$ by imposing $\eta\geq 0$ and $\nu> 0$. The solution (\ref{RR}) for the metric function $R(t)$, indicates that for a fixed number of dimension, the value of $R(t)$ increases by increasing the norm of the coupling constant $|a|$.

\begin{figure}[H]
	\centering
		\begin{subfigure}{0.4\linewidth}
		\includegraphics[width=\linewidth]{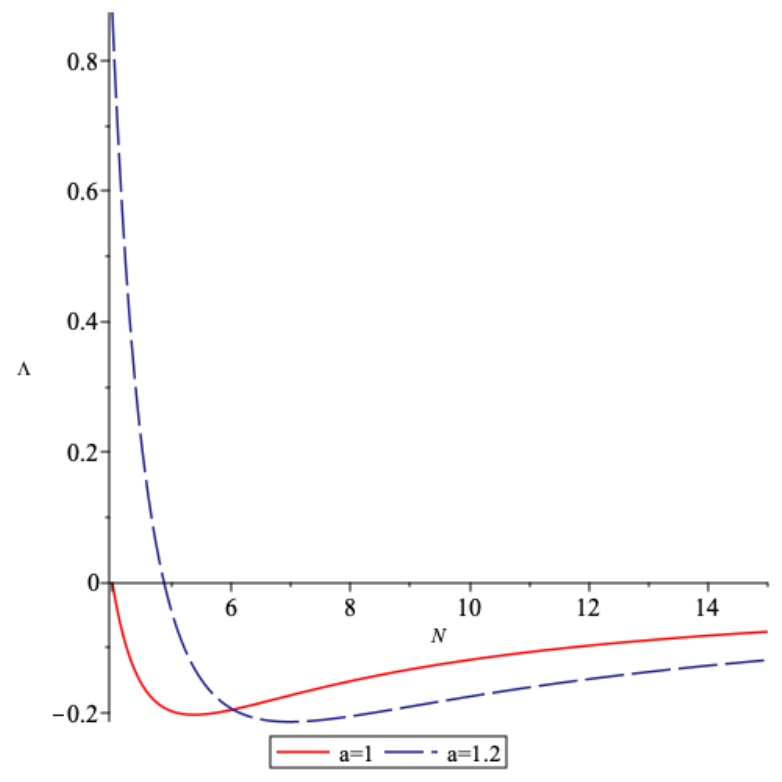}
		\caption{\centering}
		\label{fig:2c}
	\end{subfigure}
	\begin{subfigure}{0.4\linewidth}
		\includegraphics[width=\linewidth]{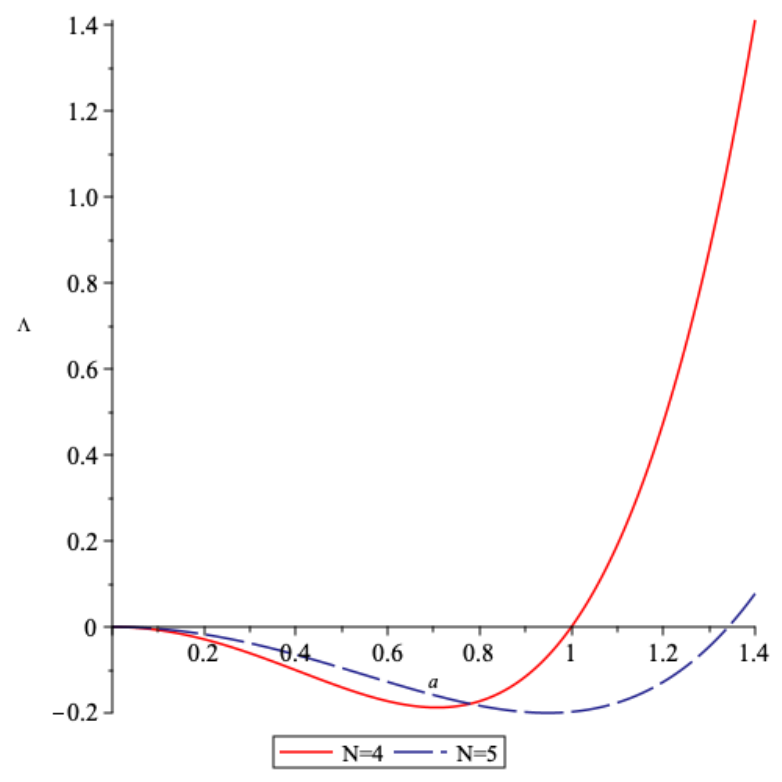}
		\caption{\centering} 
		\label{fig:2d}
	\end{subfigure}
	\caption{ The behaviour of the cosmological constant in terms of (figure a) the spatial dimension $N$ for two different values of the coupling constant $a$, and (figure b) in terms of $N$ for two different values of $a$, where we set $\eta=1$}
	\label{fig:Lamb}
\end{figure}

\begin{figure}[h]
    \centering
    \includegraphics[scale=0.4]{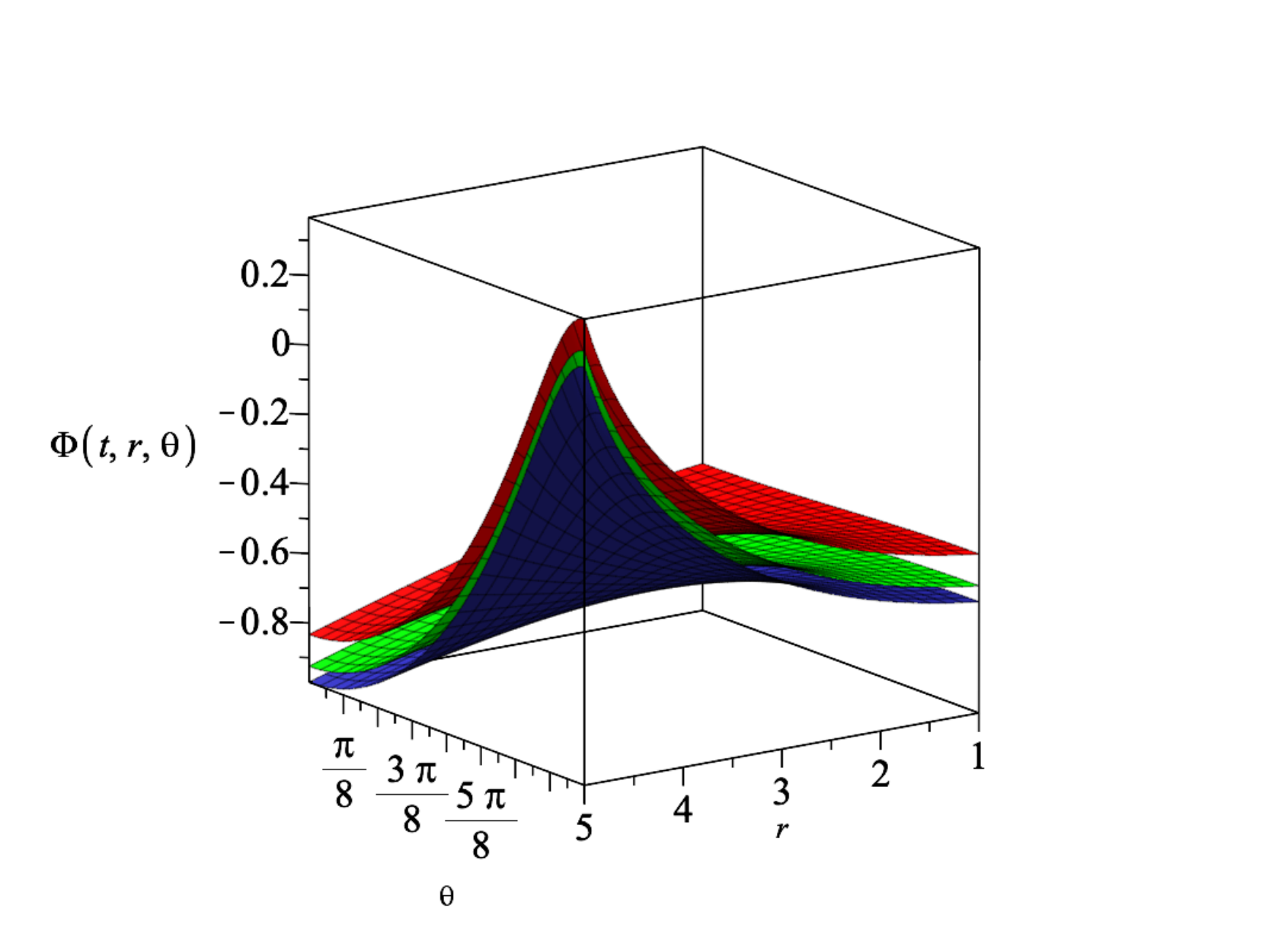}
    \caption{The behaviour of the dilaton field $\Phi(t,r,\theta)$ for three different spatial dimensions $N=4$, $N=5$ and $N=6$, which correspond to the upper, middle and lower surface, respectively. We set the constants $g_+=0.5$, $g_-=15$, $\eta=1$, $\nu=2$ and $a=1$, and assume the time slice $t=1$. }
    \label{figure:Phi}
\end{figure}

From equation (\ref{cosmoc}), we realize that depending on the value of the coupling constant $a$, and the number of spatial dimension $N$, the cosmological constant $\Lambda$ can become positive, negative or zero. For example, in (4+1)-dimensions, $\Lambda>0$ (dS space) when the coupling constant $a>1$, $\Lambda=0$ for $a=\pm 1$, and $\Lambda<0$ (AdS space) for $-1<a<1$. As an example, in figure \ref{fig:Lamb}, we illustrate the changes in the cosmological constant $\Lambda$ with respect to the spatial dimensions $N$ for different values of the coupling constant $a$, and its changes with respect to the coupling constant $a$ for different values of $N$.
We find the following relation between the coupling constants $a$ and $b$ by solving the other non-zero components of the field equations
\begin{equation}
    ab=-(N-2), \label{const}
\end{equation}
which indicates that the coupling constants $a$ and
$b$ cannot be equal to each other. Moreover, we can rewrite the action of Einstein-Maxwell-dilaton theory (\ref{action}), as
\begin{equation}
    S=\int d^{N+1}x\sqrt{-g}\{R-\frac{4}{N-1}(\nabla\phi)^2-e^{-4/(N-1)a\phi}F^2-e^{-\frac{4(N-2)}{a(N-1)}\phi}\Lambda\},
\end{equation}
from which we realize that increasing the coupling constant $a$, leads to an increase in the strength of the interaction between the dilaton field and the cosmological constant, and the decrease in the strength of the interaction between the dilaton field and the electromagnetic field. It is worth noting that our $N+1$-dimensional results are all independent of the constants $k$ and $c$ that appear in the Bianchi type IX geometry (\ref{trBIX}). Therefore, our results satisfy all the field equations for any sub-classes of the Bianchi type IX geometry. 
Substituting the result for the metric functions $H(r,\theta)$ and $R(t)$ into the dilaton field $\Phi(t,r,\theta)$ (\ref{dil}), we get
\begin{equation}
    \Phi(t,r,\theta)=\frac{\left(N -1\right) \left(\left(-N +2\right) \ln \! \left(\mathit{g_+} \,r^{2} \cos \! \left(\theta \right)+\mathit{g_-} \right)+\ln \! \left(\mathit{\eta} t +\mathit{\nu} \right) \left(a^{2}+N -2\right)\right) a}{2 \left(N -2\right) \left(a^{2}+N -2\right)}.\label{dilfin1}
\end{equation}
We show the behaviour of the dilaton field $\Phi(t,r,\theta)$ for three different spatial dimensions in figure \ref{figure:Phi}. We notice that for a specific time slice, the dilaton field decreases by increasing the dimension.

In order to study the singularities of this spacetime, we calculate the Ricci scalar and the Kretschmann invariant of the solutions in $N+1$-dimensions
\begin{equation}
R_N=\frac{f^{(N)}(t,r,\theta,\psi)}{r^{10}R(t)^2H(r,\theta)\sin^2{\theta}},  
\end{equation}
\begin{equation}
    \mathcal{K}_N=\frac{g^{(N)}(t,r,\theta,\psi)}{r^{14}R(t)^2H(r,\theta)\sin^2{\theta}},
\end{equation}
where we show the numerator of these expressions by $f^{(N)}(t,r,\theta,\psi)$ and $g^{(N)}(t,r,\theta,\psi)$, respectively, which are functions of the coordinates. These functions change based on the number of dimensions. We realize that in any dimension $N\geq 4$, the Ricci scalar and the Kretschmann invariant diverge at $r=0$, $\sin{\theta}=0$ and on the hypersurface $H(r,\theta)=0$. We can avoid the singularities  at $R(t)=0$, by restricting the constants $\eta \geq 0$ and $\nu>0$.

\begin{figure}[h]
    \centering
    \includegraphics[scale=0.5]{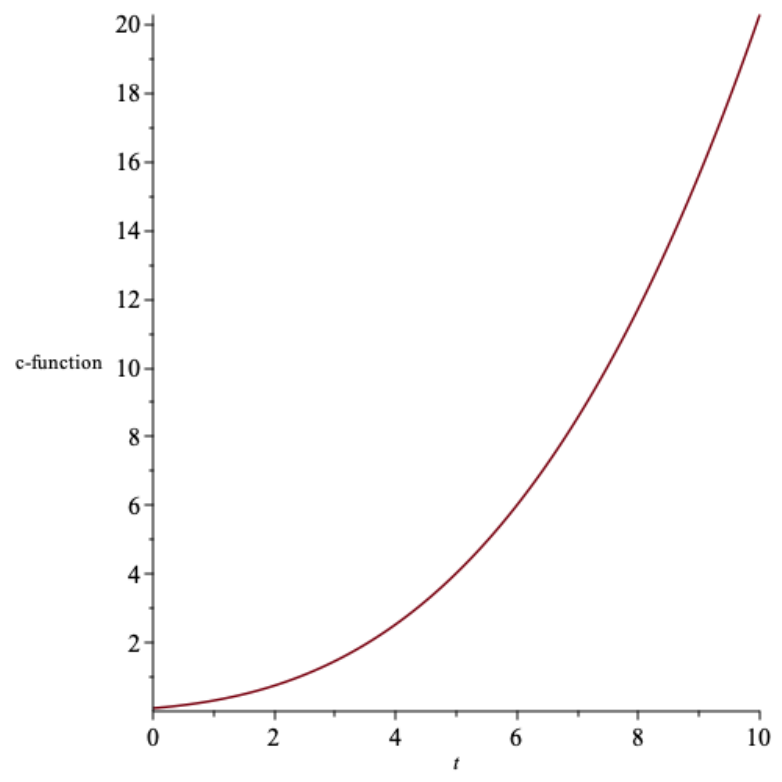}
    \caption{The behaviour of the c-function with respect to the time coordinate in $N=4$, for a set of values for the constants. }
    \label{figure:cab}
\end{figure}

There is a correspondence in asymptotically AdS/dS spacetimes, between the UV/IR physics in dual conformal field theory and the near boundary of the spacetime. The $c$-theorem states that for an expanding dS spacetime, the renormalization group flows to the ultraviolet, and for a contracting dS spacetime, to the infrared \cite{ghezelbash2010cosmological}-\cite{balasubramanian2002mass}. The $c$-function in $N+1$-dimensions is given by
\begin{equation}
    c\sim \frac{1}{(G_{tt})^{\frac{N-1}{2}}},
\end{equation}
where $G_{tt}$ is the effective Einstein field tensor. In figure \ref{figure:cab}, we show the behaviour of the c-function in $N=4$.

\begin{figure}[h]
	\centering
		\begin{subfigure}{0.49\linewidth}
		\includegraphics[width=\linewidth]{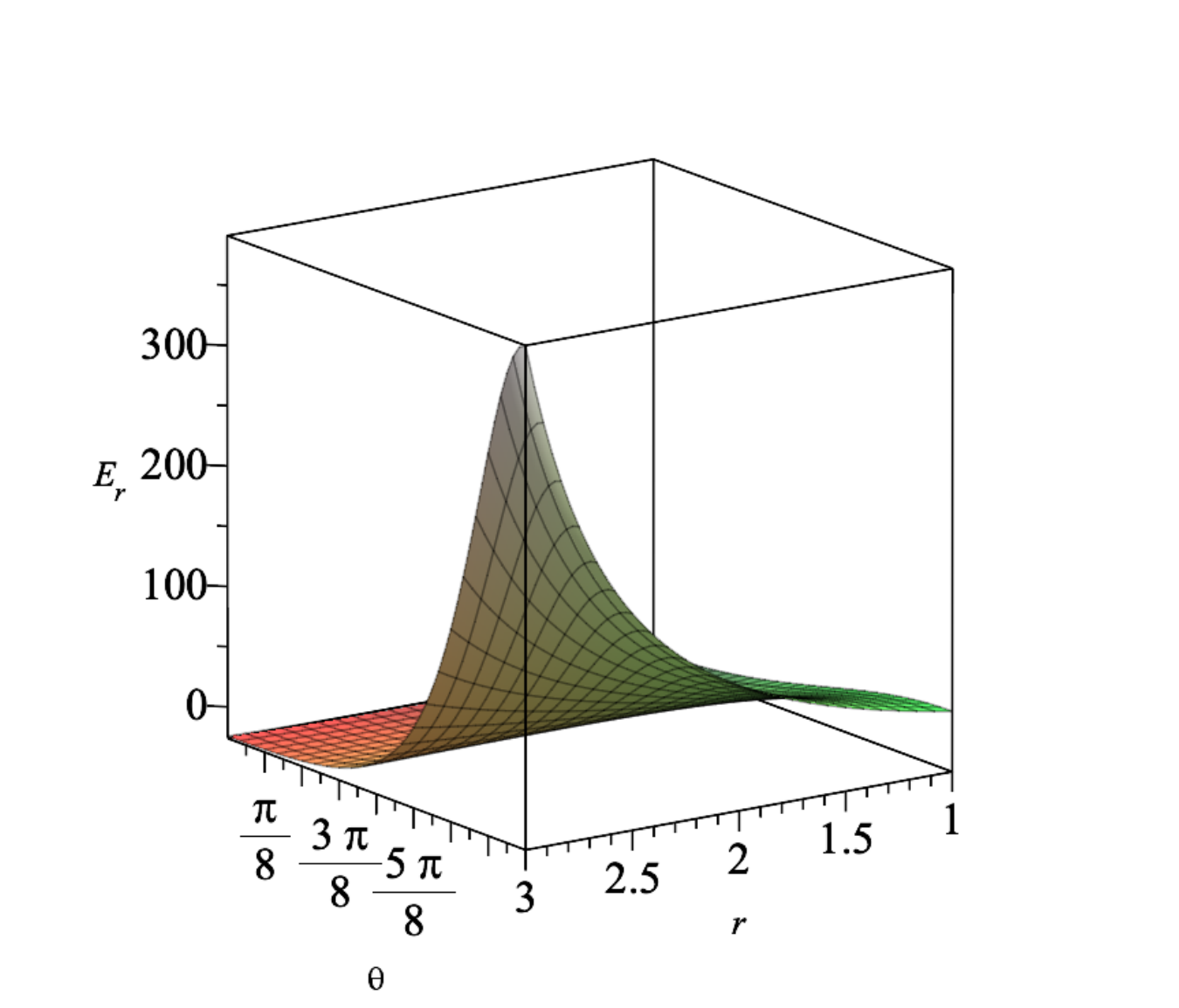}
		\caption{\centering}
	\end{subfigure}
	\begin{subfigure}{0.49\linewidth}
		\includegraphics[width=\linewidth]{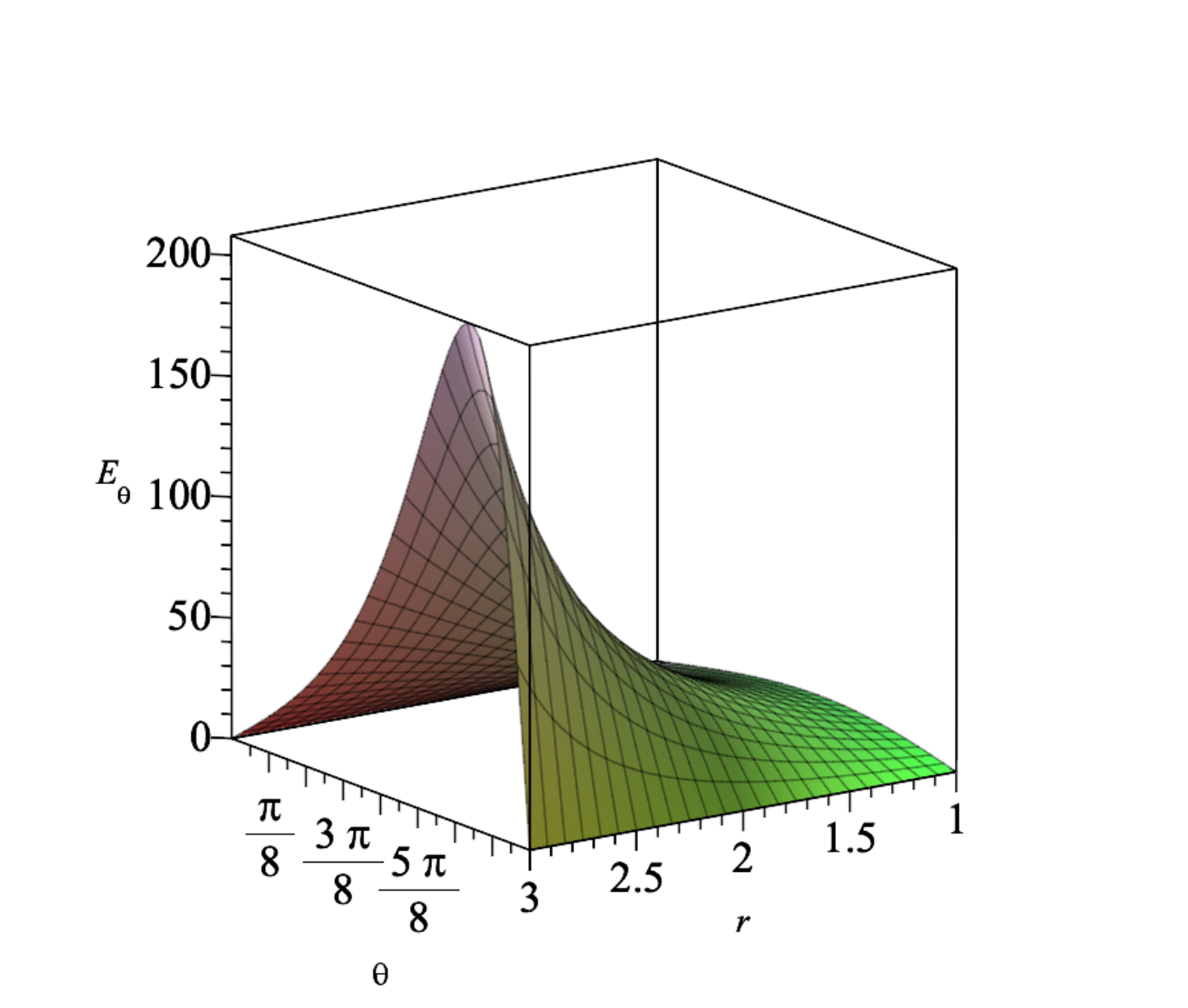}
		\caption{\centering} 
	\end{subfigure}
	\caption{ The electric fields (figure a): $E_r$ and (figure b): $E_{\theta}$ in $(4+1)$-dimensions at $t=5$, where we set $a=3$, $\eta=1$, $\nu=2$, $g_+=1$ and $g_-=15$. }
	\label{fig:Eneq}
\end{figure}

The electromagnetic gauge field $A_{\mu}$ given in equation (\ref{A}) is explicitly given by
\begin{equation}
A_t(t,r,\theta)=\alpha(\eta t+\nu)^{\frac{a^2}{N-2}}(g_+r^2\cos\theta+g_-)^{-1}\label{AAE},
\end{equation}
which yields the electric field in $r$ and $\theta$ directions. Furnished with (\ref{AAE}), we find the components of the electric field are given by
\begin{equation}
    E_{r}=-\frac{r\cos{\theta}g_+(\eta t+\nu)^{\frac{a^2}{N-2}}\sqrt{2N-2}}{(g_+r^2\cos{\theta}+g_-)^2\sqrt{a^2+N-2}},
\end{equation}
\begin{equation}
    E_{\theta}=\frac{r^2\sin{\theta}g_+(\eta t+\nu)^{\frac{a^2}{N-2}}\sqrt{2N-2}}{2(g_+r^2\cos{\theta}+g_-)^2\sqrt{a^2+N-2}}.
\end{equation}
We show the behaviour of these electric fields in figure \ref{fig:Eneq} for assumed values for the constants.


We show in section \ref{sec:up} that our solutions in this section can be uplifted to a higher dimensional Einstein-form theory, however they can't be uplifted into a higher dimensional Einstein-Maxwell theory with cosmological constant, or just higher dimensional Einstein gravity with a cosmological constant. In fact, in the first uplifting situation, the uplifting works if $a=b$, and the number of extra spatial directions is given by \cite{gouteraux2012holography,gouteraux2011generalized}
\begin{equation}
d=\frac{3a^2}{1-a^2}.
\end{equation}
However as we noticed before, the consistency of the solutions implies $ab=-(N-2)$ and so, there is no value for the coupling constant $a$, such that $a=b$. The latter uplifting, i.e. the case of uplifting the $N+1$-dimensional solutions to Einstein-Maxwell-dilaton theory with two coupling constants to the solutions of $N+2$-dimensional gravity with the cosmological constant, is only possible if the coupling constants are equal to $a=\pm 2$ and $b=\pm \frac{1}{2}$, \cite{ch1}, \cite{ch2}.  However, the product of these coupling constants are equal to $+1$, and so they are not in agreement with the relation between the coupling constants (\ref{const}). We actually explicitly check out that the $N+2$-dimensional 
metric
\begin{equation}
ds_{N+2}^2=e^{\mp(\frac{4}{3})(\frac{1}{2})\phi(t,r,\theta)}ds_{N+1}^2+e^{\pm4(\frac{1}{2})\phi(t,r,\theta)}(dz+2A_t(t,r,\theta)dt)^2,\label{upl}
\end{equation}
where $ds_{N+1}^2$, $\phi(t,r,\theta)$ and $A_t(t,r,\theta)$ are given by (\ref{st}), (\ref{dilfin1}), and (\ref{AAE}), respectively, 
does not satisfy the field equations of the $N+2$-dimensional gravity with a cosmological constant. We note that in (\ref{upl}), $z$ is the uplifted coordinate. In fact,  a careful analysis of the uplifting process in \cite{ch2}, shows that the $N+1$-dimensional metric for the Einstein-Maxwell-dilaton theory, always is diagonal; though our metric ansatz (\ref{st}) has off-diagonal elements due to the Bianchi type IX geometry. It looks like that 
we may need 
a new anstaz for the uplifting of the $N+1$-dimensional Einstein-Maxwell-dilaton theory into a $N+2$-dimensional gravity with the cosmological constant.

In the next section, we propose a different set of ansatzes to find a new exact solution to the theory, which includes the case where the coupling constants are equal to each other $a=b$.

\section{Einstein-Maxwell-dilaton theory with two non-zero Equal coupling constants}
\label{sec:equal}
One of the main results of the previous section is the restriction on the coupling constants $ab=-(N-2)$, which indicates that with the considered ansatzes (\ref{st}), (\ref{dil}), and (\ref{A}), the coupling constants can't be equal to each other. In order to explore a new class of exact solution to the Einstein-Maxwell-dilaton theory with equal coupling constants $a=b$, we assume a new ansatz for the $N+1$-dimensional spacetime as
\begin{equation}
    ds_{N+1}^2=-\frac{1}{H(t,r,\theta)^2}dt^2+H(t,r,\theta)^{\frac{2}{(N-2)}}R(t)^2[ds_{B. IX}^2+\Sigma_{i=1}^{N-4}dx^2_i], \label{st2}
\end{equation}
where the metric function $H(t,r,\theta)$ is now a function of the spatial coordinates $r$ and $\theta$, as well as the time coordinate $t$. We also consider a new ansatz for the electromagnetic gauge
\begin{equation}
  A_t(t,r,\theta)=\alpha R^X(t)H^Y(t,r,\theta), \label{Meq}
\end{equation}
and the dilaton field
\begin{equation}
    \phi(t,r,\theta)=-\frac{(N-1)}{4a}\ln{(H^U(t,r,\theta)R^V(t))}, \label{Deq}
\end{equation}
where $X$, $Y$, $U$ and $V$ are constants that will be determined by the field equations. 

\begin{figure}[h]
	\centering
		\begin{subfigure}{0.49\linewidth}
		\includegraphics[width=\linewidth]{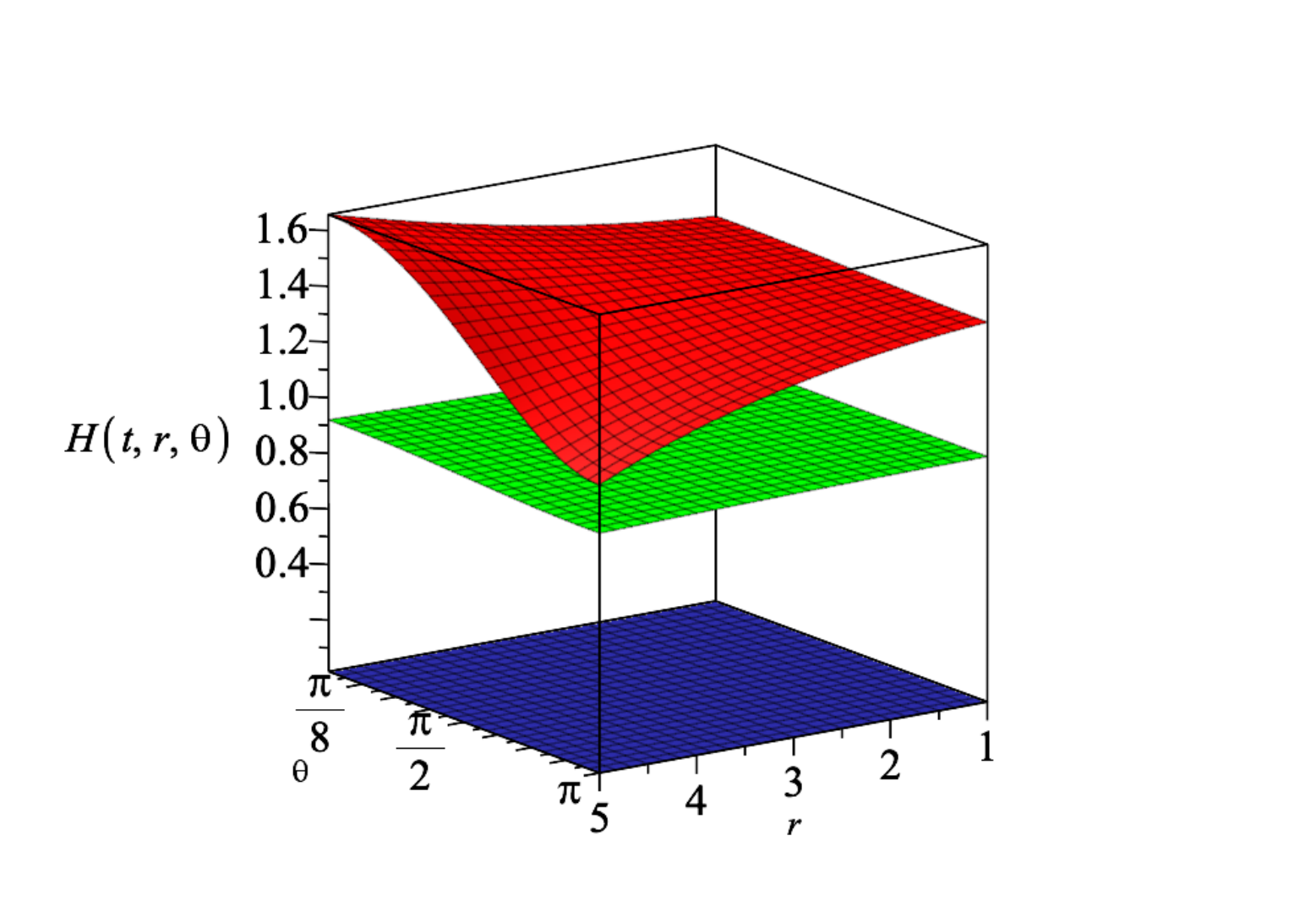}
		\caption{\centering}
	\end{subfigure}
	\begin{subfigure}{0.49\linewidth}
		\includegraphics[width=\linewidth]{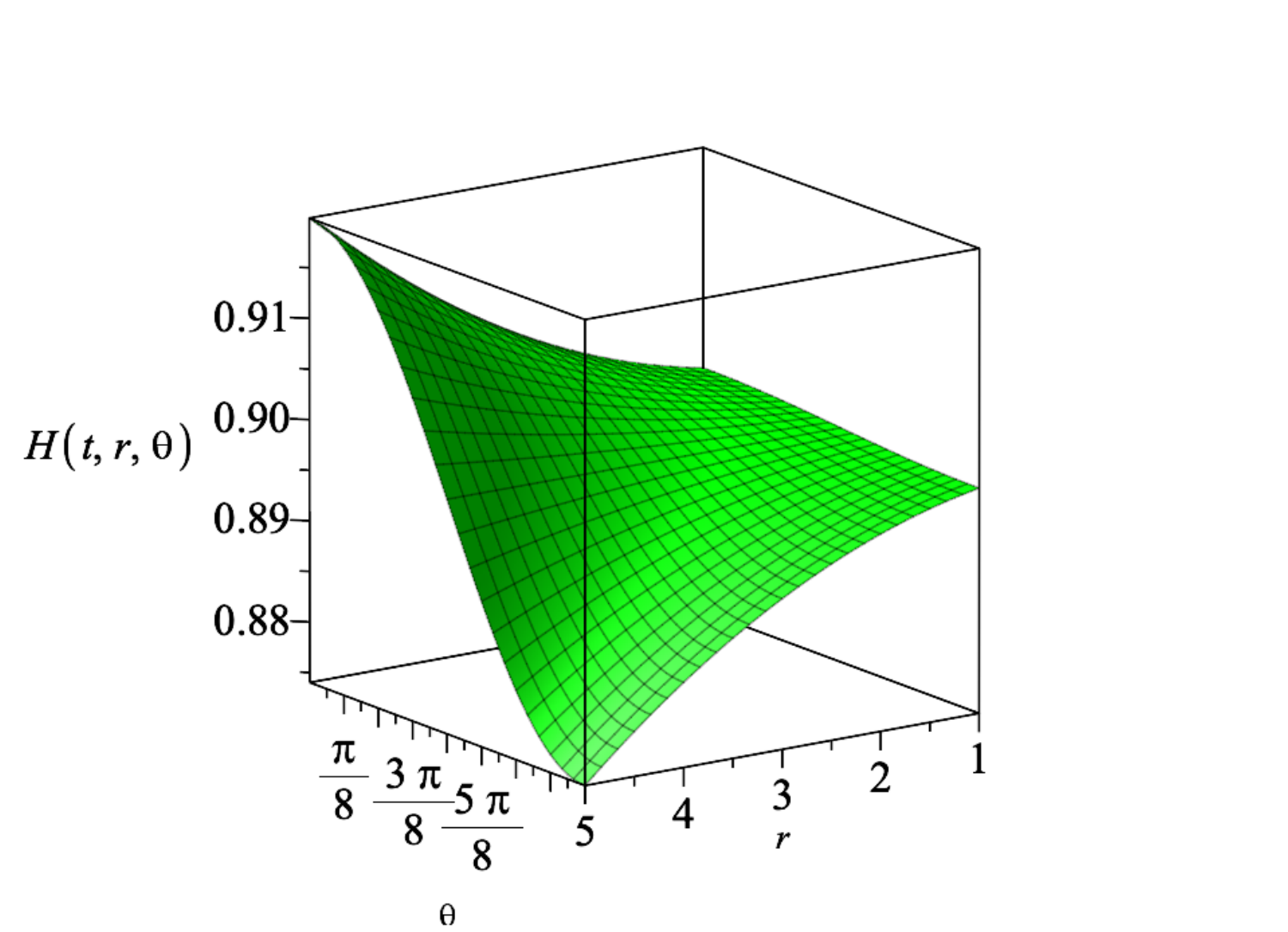}
		\caption{\centering} 
	\end{subfigure}
	    \caption{The metric function $H(t,r,\theta)$  in terms of the coordinates $r$ and $\theta$ for three different spatial dimensions $N=4$, $N=5$ and $N=6$ (figure a), which correspond to the upper, middle and lower surface, respectively. The metric function $H(t,r,\theta)$ in terms of the coordinates $r$ and $\theta$ for $N=5$ (figure b), which is the zoom in of middle surface in figure a. We assumed the constants $\eta=1$, $\nu=2$, $g_+=0.5$, $g_-=5$, $a=1.5$ and the time slice $t=2$. }
	\label{fig:Haa}
\end{figure}

Combining the electromagnetic field equations $\mathcal{M}_r$ and $\mathcal{M}_{t}$, leads to a differential equation for the metric function $H(t,r,\theta)$, which can be solved as
\begin{equation}
    H(t,r,\theta)=R^{-(N-2)}(t)\Bigl(R(t)^p+g_+r^2\cos{\theta}+g_- \Bigl)^{\frac{N-2}{N-2+a^2}}, \label{Hequal}
\end{equation}
where $g_{\pm}$ and $p$ are arbitrary constants. Substituting the result for $H(t,r,\theta)$ in the other non-zero components of the electromagnetic and Einstein field equations, we determine the constants $X$, $Y$ and $\alpha$ in the Maxwell gauge field (\ref{Meq}) 
\begin{equation}
    X=-a^2, Y=-1-\frac{a^2}{(N-2)}, \alpha^2=\frac{(N-1)}{2(a^2+(N-2))},
\end{equation}
and the constants $U$ and $V$ in the dilaton field (\ref{Deq})
\begin{equation}
    U=\frac{2a^2}{(N-2)}, V=2a^2.
\end{equation}
In figure \ref{fig:Haa}, we represent the behaviour of the metric function $H(t,r,\theta)$ with respect to the coordinates $r$ and $\theta$, for three different dimensions, where we set the coupling constant $a=1.5$. As we notice from figure \ref{fig:Haa}a, the fluctuations in the metric function is not visible especially for $N=5$ and $N=6$, because of the vertical axis values. In figure \ref{fig:Haa}b, we zoom in on the metric function for $N=5$, and notice non-trivial fluctuations versus the coordinates $r$ and $\theta$. We notice that for a specific time slice, the metric function decreases by increasing the dimension of spacetime.
 
Substituting these results in the Einstein equation $\mathcal{G}_{tt}$ and $\mathcal{G}_{rr}$, we find the metric function $R(t)$ and the cosmological constant as
\begin{equation}
    R(t)=(\eta t+\nu)^{\Delta},
\end{equation}
\begin{equation}
    \Lambda=-(N-1)\frac{\eta^2(a^2-N)}{(a^2+N-2)^2}(\frac{a^2+2}{a^2+4-N})^2, \label{lameq}
\end{equation}
where $\eta$ and $\nu$ are arbitrary constants, and $\Delta$ is a constant that has the following relation with $p$ that appears as the power of $R(t)$ in equation (\ref{Hequal}) 
\begin{equation}
    p=\frac{(N-2)\Delta+1}{\Delta}.
\end{equation}

In order to reproduce the same results of the reference \cite{fahim2021new} for the five-dimensional Einstein-Maxwell-dilaton theory, based on the Bianchi type IX geometry, we consider $p=a^2+2$ with no loss of generality. This choice makes $\Delta=\frac{1}{a^2+4-N}$. We note that the divergence in the $\Delta$ can be removed by choosing different values for $p$.

\begin{figure}[h]
	\centering
		\begin{subfigure}{0.49\linewidth}
		\includegraphics[width=\linewidth]{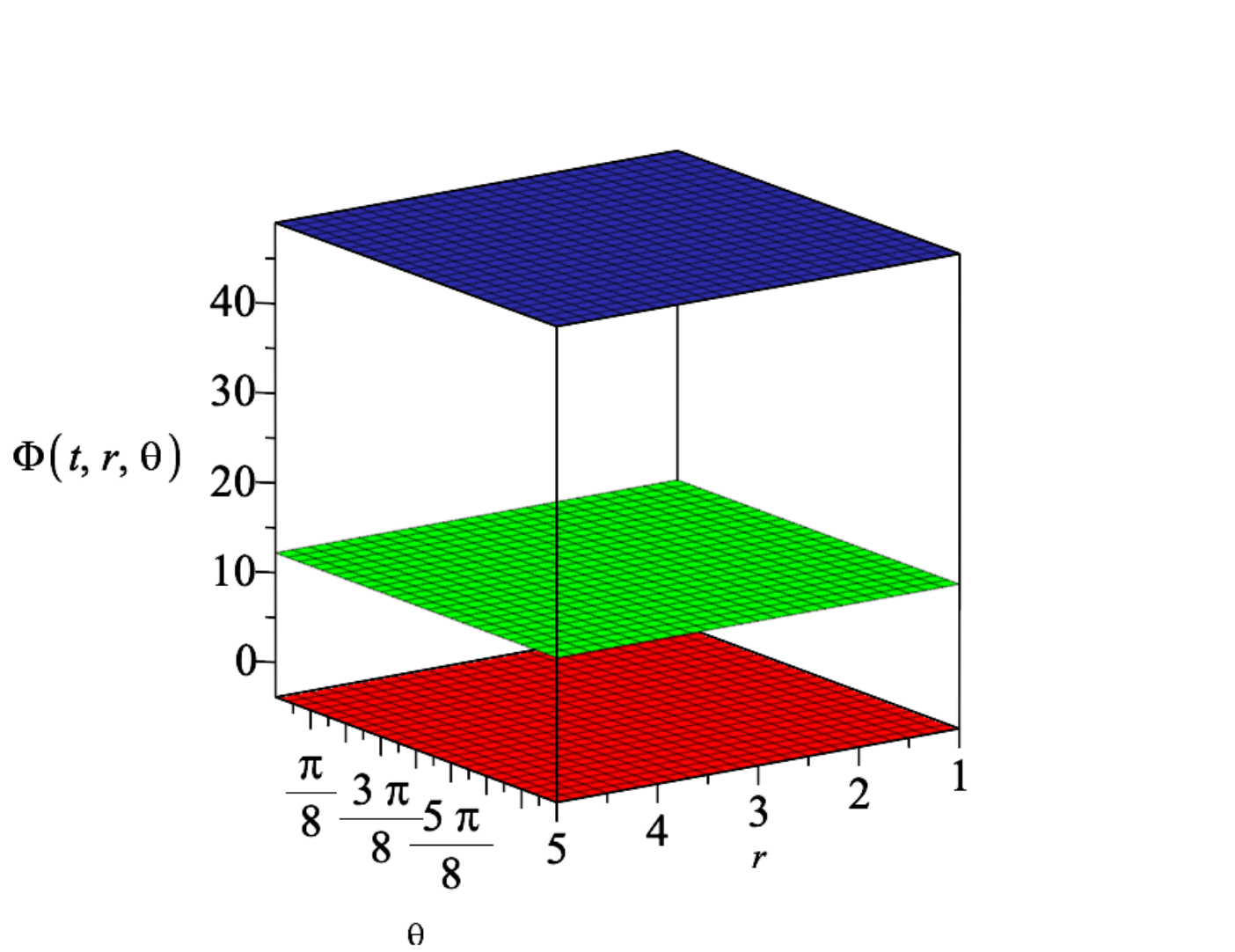}
		\caption{\centering}
	\end{subfigure}
	\begin{subfigure}{0.49\linewidth}
		\includegraphics[width=\linewidth]{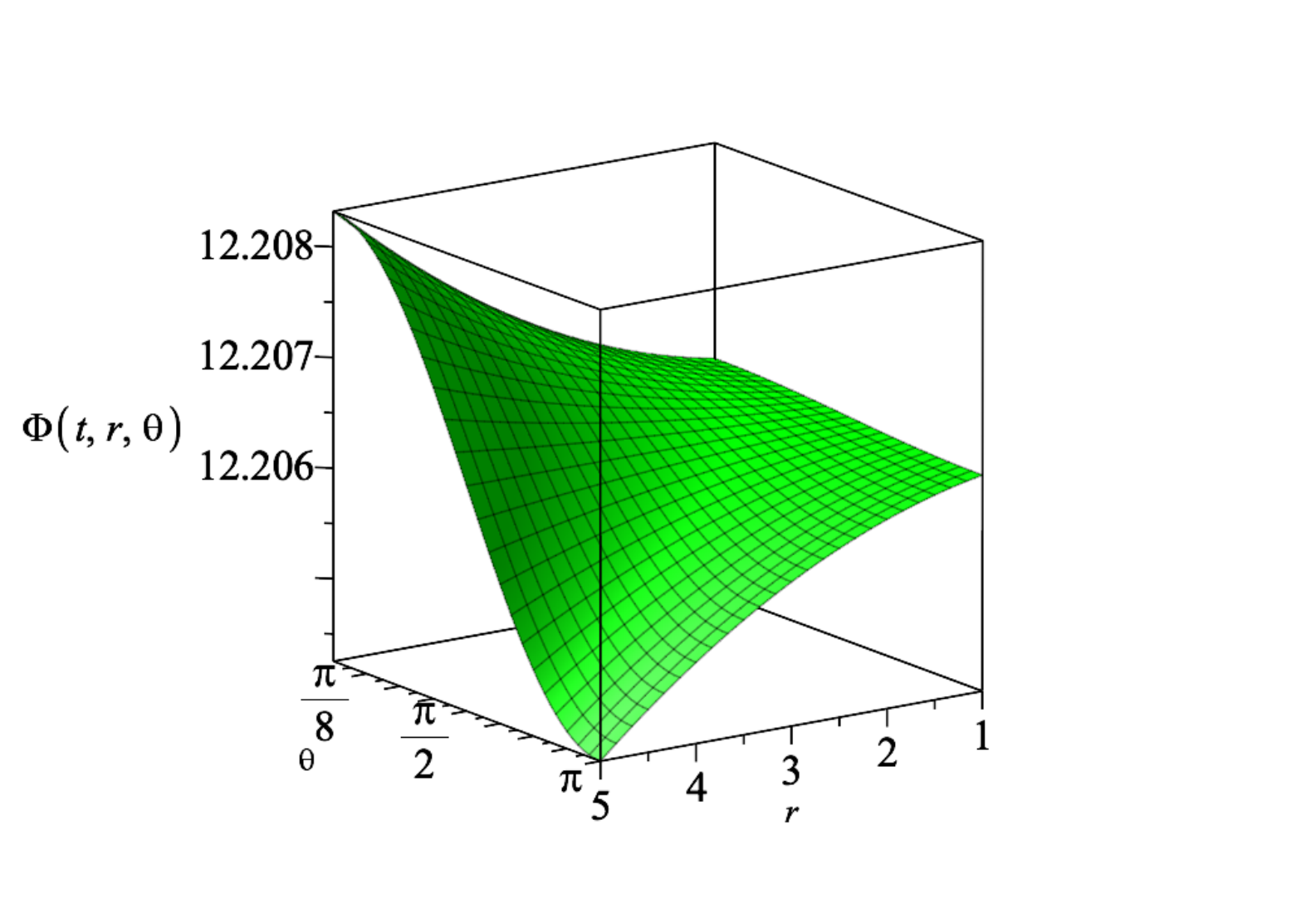}
		\caption{\centering} 
	\end{subfigure}
	    \caption{The dilaton field $\Phi(t,r,\theta)$  in terms of the coordinates $r$ and $\theta$ for three different spatial dimensions $N=4$, $N=5$ and $N=6$ (figure a), which correspond to the lower,  middle and upper surface, respectively. The dilaton field $\Phi(t,r,\theta)$ in terms of the coordinates $r$ and $\theta$ for $N=5$ (figure b), which is the zoom in of middle surface in figure a. We assumed the constants $\eta=1$, $\nu=2$, $g_+=0.5$, $g_-=5$, $a=1.5$ and the time slice $t=2$. }
	\label{fig:Paa}
\end{figure}

Substituting the results for the metric functions $H(r,\theta)$ and $R(t)$, into the dilaton field $\Phi(t,r,\theta)$ (\ref{Deq}), we get
\begin{equation}
    \Phi(t,r,\theta)=-\frac{2a(N-1)}{N-2+a^2}\ln\{{(\eta t+\nu)^{a^2+2}}+g_+r^2\cos\theta+g_-\}\label{dilfin}.
\end{equation}
We show the behaviour of the dilaton field $\Phi(t,r,\theta)$ for three different spatial dimensions in figure \ref{fig:Paa}. 
As we notice from figure \ref{fig:Paa}a, the fluctuations in the dilaton field is not visible for different values of $N=4$, $N=5$ and $N=6$, because of the vertical axis values. In figure \ref{fig:Paa}b, we zoom in on the dilaton field for $N=5$, and notice non-trivial fluctuations versus the coordinates $r$ and $\theta$. 
We notice that for a specific time slice, the dilaton field increases by increasing the dimension of spacetime.

\begin{figure}[H]
	\centering
		\begin{subfigure}{0.4\linewidth}
		\includegraphics[width=\linewidth]{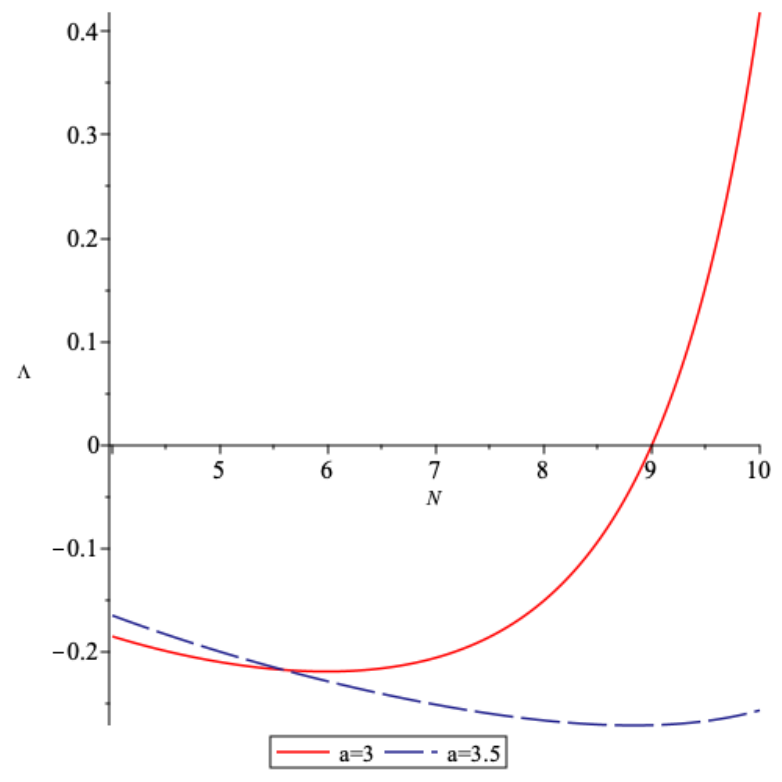}
		\caption{\centering}
		\label{fig:2c}
	\end{subfigure}
	\begin{subfigure}{0.4\linewidth}
		\includegraphics[width=\linewidth]{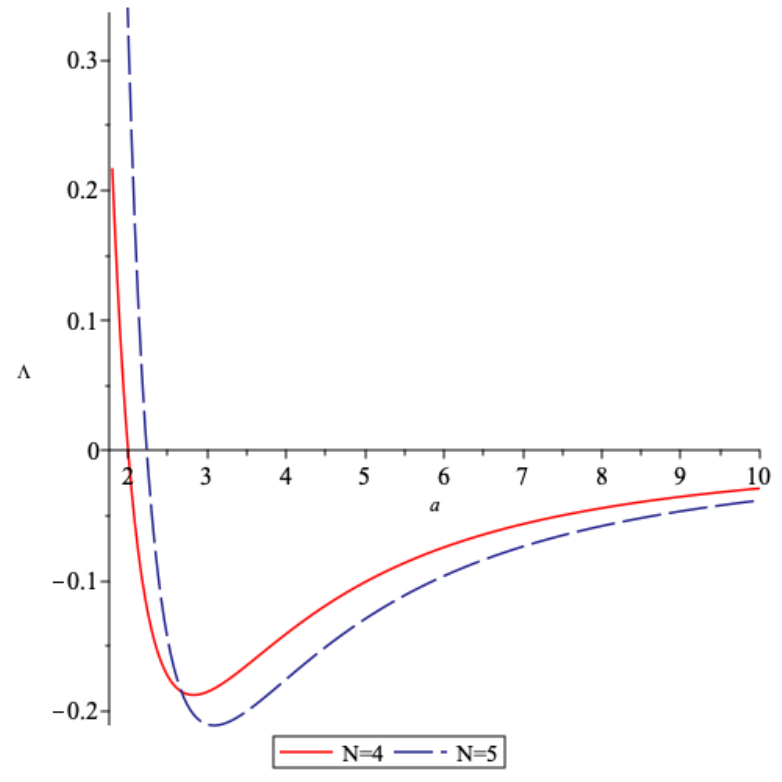}
		\caption{\centering} 
		\label{fig:2d}
	\end{subfigure}
	\caption{ The behaviour of the cosmological constant $\Lambda$ in terms of (figure a): the number of spatial dimension $N$ for two different values of the coupling constant $a$, and (figure b): in terms of $a$ in two different dimensions,  where we assumed $\eta=1$.}
	\label{fig:Lequ}
\end{figure}

We realize that the cosmological constant $\Lambda$ in $N+1$-dimensions can be positive, negative or zero, based on the coupling constant $a$ and the number of the spatial dimension $N$. In figure \ref{fig:Lequ}, we show the behaviour of the cosmological constant $\Lambda$ in terms of the number of dimension $N$ in figure \ref{fig:Lequ}a , and in terms of the coupling constant $a$,  for different values of $N$, in figure \ref{fig:Lequ}b.

It's worth mentioning that for the case where the coupling constant $a=1$, in $N=4$ dimensions, the action of Einstein-Maxwell-dilaton theory reduces to the low-energy effective action for heterotic string theory \cite{rocha2018self}.

\begin{figure}[H]
    \centering
    \includegraphics[scale=0.50]{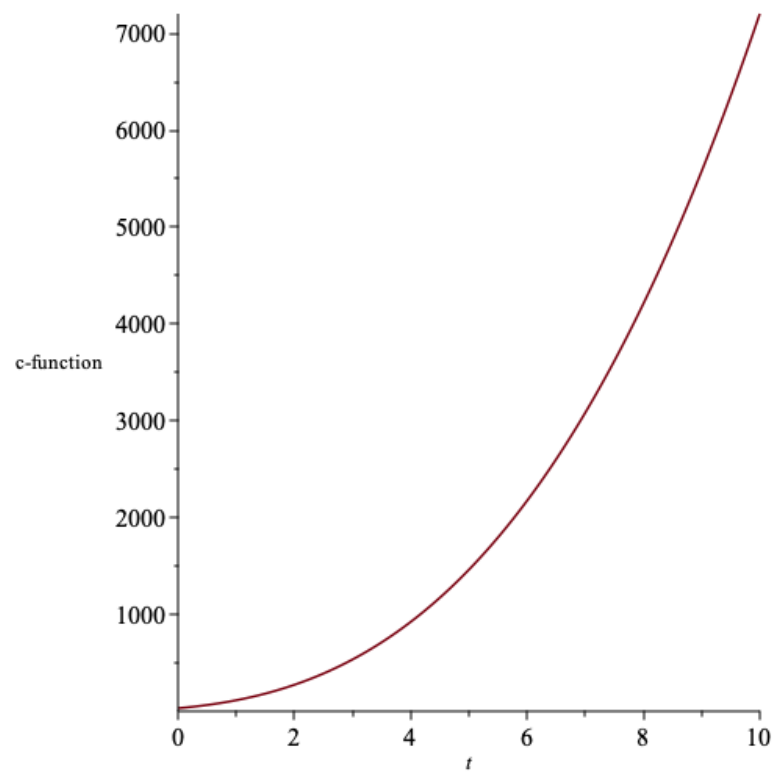}
    \caption{The behaviour of the c-function for $N=4$ for a set of values for the constants. }
    \label{figure:ceq}
\end{figure}

In figure \ref{figure:ceq}, we represent the c-function of the solutions, where $N=4$.

The electromagnetic gauge field $A_{\mu}$ given in equation (\ref{Meq}), is explicitly given by
\begin{equation}
A_t(t,r,\theta)=\alpha(\eta t+\nu)^{-\frac{a^2}{a^2+4-N}}\{(\eta t+\nu)^{\frac{N-2+a^2}{a^2+4-N}}((\eta t+\nu)^{\frac{2+a^2}{a^2+4-N}}+g_+r^2\cos\theta+g_-)^{-1}\}\label{AAE2},
\end{equation}
which yields the electric field in $r$ and $\theta$ direction. Furnished with (\ref{AAE2}), we find the components of the electric field are given by
\begin{equation}
    E_{r}=-\frac{r\cos{\theta}g_+(\eta t+\nu)^{\frac{2a^4-N+2}{-a^2+N-4}}\sqrt{2N-2}}{\Bigl(1+(g_+r^2\cos{\theta}+g_-)(\eta t+\nu)^{\frac{a^4}{-a^2+N-4}}\Bigl)^2\sqrt{a^2+N-2}},
\end{equation}
\begin{equation}
    E_{\theta}=\frac{r^2\sin{\theta}g_+(\eta t+\nu)^{\frac{2a^4-N+2}{-a^2+N-4}}\sqrt{2N-2}}{\Bigl(1+(g_+r^2\cos{\theta}+g_-)(\eta t+\nu)^{\frac{a^4}{-a^2+N-4}}\Bigl)^22\sqrt{a^2+N-2}}.
\end{equation}

\begin{figure}[h]
	\centering
		\begin{subfigure}{0.49\linewidth}
		\includegraphics[width=\linewidth]{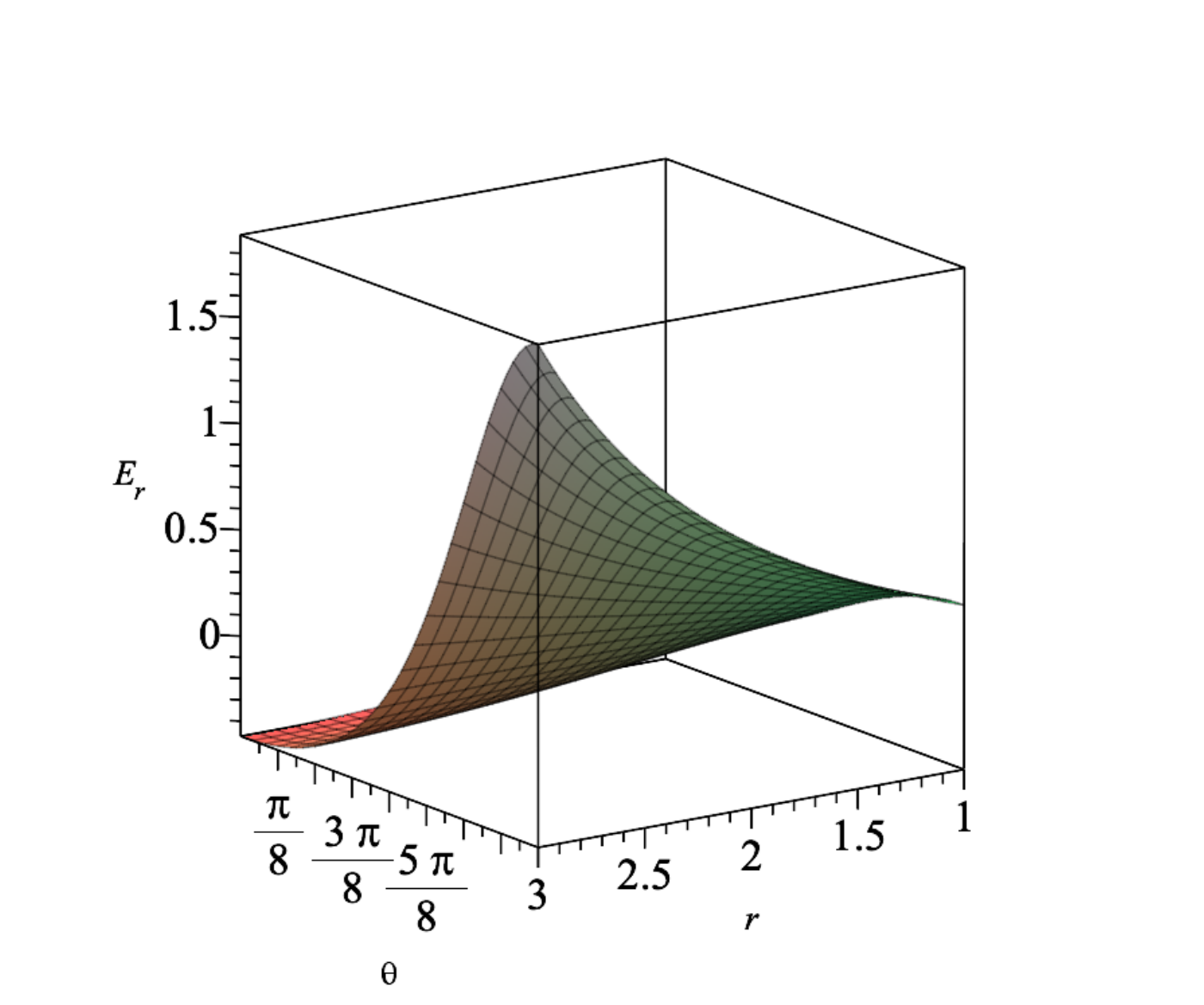}
		\caption{\centering}
	\end{subfigure}
	\begin{subfigure}{0.49\linewidth}
		\includegraphics[width=\linewidth]{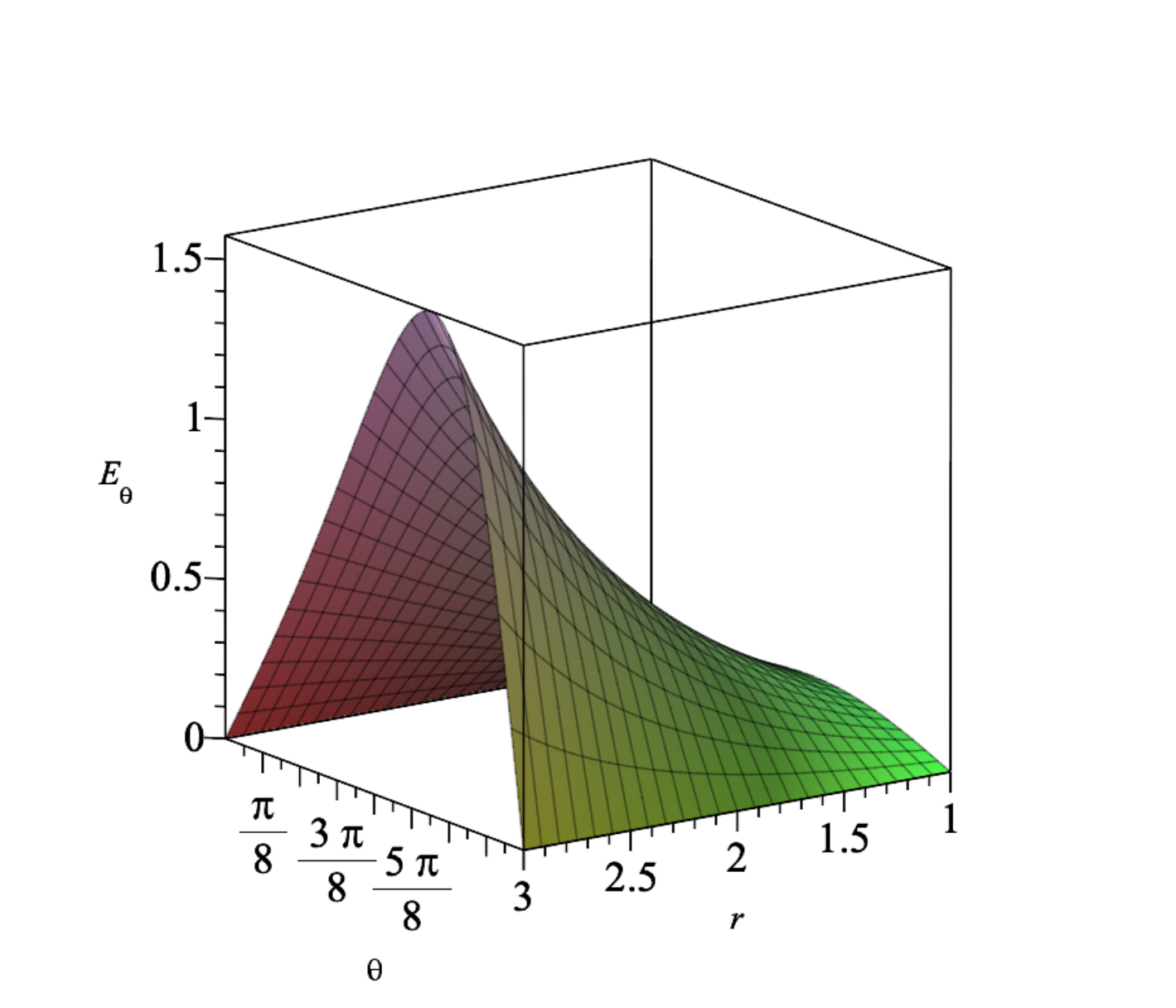}
		\caption{\centering} 
	\end{subfigure}
	\caption{ The electric fields (figure a): $E_r$ and (figure b): $E_{\theta}$ in $(4+1)$-dimensions for $a=b=1$, where we set $t=10$, $\eta=1$, $\nu=2$ , $g_+=1$ and $g_-=15$.}
	\label{fig:Eeq}
\end{figure}

In figure \ref{fig:Eeq}, we show the electric fields $E_r$ and $E_{\theta}$ in terms of the coordinates $r$ and $\theta$, for a set values for the constants.

We also note that the $N+1$-dimensional metric (\ref{st2})
can be uplifted to the solution to the higher dimensional Einstein-Maxwell theory with a cosmological constant, if the coupling constant $a$ is greater than or equal to  $\frac{1}{2}$ and less than $1$ \cite{ch2}.  The Einstein-Maxwell theory with a cosmological constant is a $N+1+{\cal D}$-dimensional theory with the cosmological constant equal to $\Lambda=\frac{9\mu^2}{2}\frac{({\cal D}+3)({\cal D}+4)}{{\cal D}^2}$, where ${\cal D}=\frac{3a^2}{1-a^2}$. The $N+1+{\cal D}$-dimensional metric is given by
\begin{equation}
ds_{N+1+{\cal D}}^2=e^{\frac{4}{3}\sqrt\frac{{\cal D}}{{\cal D}+3}\phi(t,r,\theta)}ds_{N+1}^2+e^{\frac{-4}{{\cal D}}\sqrt\frac{{\cal D}}{{\cal D}+3}\phi(t,r,\theta)}d\vec y \cdot d\vec y,\label{highdEM}
\end{equation}
where $\vec y=(y_1,\cdots ,y_{\cal D})$ parametrizes an Euclidean  $E^{\cal D}$ space, and $ds_{N+1}^2$ and $\phi(t,r,\theta)$ are given by (\ref{st2}) and (\ref{dilfin}), respectively. We explicitly check that (\ref{highdEM}) satisfies all the Einstein-Maxwell field equations for $a=\frac{1}{2},\sqrt{\frac{2}{5}}$ and $\frac{1}{\sqrt{2}}$ (where ${\cal D}=1,2$ and $3$) with the cosmological constants $\Lambda=90\mu^2,\,\frac{135}{4}\mu^2$ and $21\mu^2$, respectively. 

We should note that the solution for the dilaton field (\ref{Deq}) is not well defined, if the coupling constant $a$, is equal to zero. In other words, if we are interested in the exact solutions, in the limit where $a\rightarrow 0$, we shall impose the limit in the action (\ref{action}), and solve the corresponding field equations. In the next section, we propose a different set of ansatzes to find a new exact solution to the theory, which includes the case where the coupling constants are equal to zero $a=b=0$.

\section{Einstein-Maxwell-dilaton Theory with two zero coupling constants}
\label{sec:00}
In this section, we consider the $N+1$-dimensional Einstein-Maxwell theory in the presence of the cosmological constant. Considering $a=b=0$ in the last section makes the dilaton field (\ref{Deq}) diverges. Moreover, this assumption leads to a divergent cosmological constant  (\ref{lameq}) when $N=4$. Therefore, we assume the  following ansatz for the line element, and the electromagnetic gauge
\begin{equation}
    ds_{N+1}^2=-\frac{1}{H(t,r,\theta)^2}dt^2+H(t,r,\theta)^{\frac{2}{(N-2)}}R(t)^2[ds_{B. IX}^2+\Sigma_{i=1}^{N-4}dx^2_i], 
\end{equation}
\begin{equation}
  A_t(t,r,\theta)=\frac{\alpha}{ H(t,r,\theta)}, \label{Az}
\end{equation}
respectively. Solving the field equations, we find the metric function $H(t,r,\theta)$ as
\begin{equation}
    H(t,r,\theta)=1+(g_+r^2\cos{\theta}+g_-)R^Y(t), \label{Hz}
\end{equation}
where $g_{\pm}$ are two arbitrary constants, and $Y$ is a constant that we will determine by the field equations. Moreover, solving the $\mathcal{G}_{tt}$ component of Einstein equation, and substituting the result and the form of $H(t,r,\theta)$ (\ref{Hz}) into $\mathcal{G}_{rr}$, we find a differential equation for $R(t)$, which gives us
\begin{equation}
     R(t)=\Bigl(\eta\exp\{\epsilon(\frac{\Lambda}{X})^{1/2}t\}\Bigl)^p,
\end{equation}
where $\eta$ is an arbitrary constant, $\epsilon=\pm 1$ and $X$ and $p$ will be determined throughout the equations. From $\mathcal{G}_{r\theta}$ component of Einstein equation
\begin{equation}
    \mathcal{G}_{r\theta}=-\frac{4r^3g_+^2\sin{\theta}\cos{\theta}(\alpha^2-\frac{(N-1)}{2(N-2)})}{(g_+r^2\cos{\theta}+g_-+\exp\{(N-2)\epsilon(\frac{\Lambda}{X})^{1/2}t\})},
\end{equation}
we find the constant $\alpha^2$ that appears in electromagnetic gauge field (\ref{Az}) as
\begin{equation}
    \alpha^2=\frac{(N-1)}{2(N-2)}.
\end{equation}

Substituting these results into the rest of the equations of motion, we determine the constants $Y$ and $X$ as
\begin{equation}
    Y=-(N-2)p,  X=N(N-1)\epsilon^2.
\end{equation}
Assuming $p=1$, which makes $Y=-(N-2)$, recovers the same results proposed in \cite{fahim2021new} for the five-dimensional Einstein-Maxwell-dilaton theory, based on the Bianchi type IX geometry. 

\begin{figure}[H]
    \centering
    \includegraphics[scale=0.40]{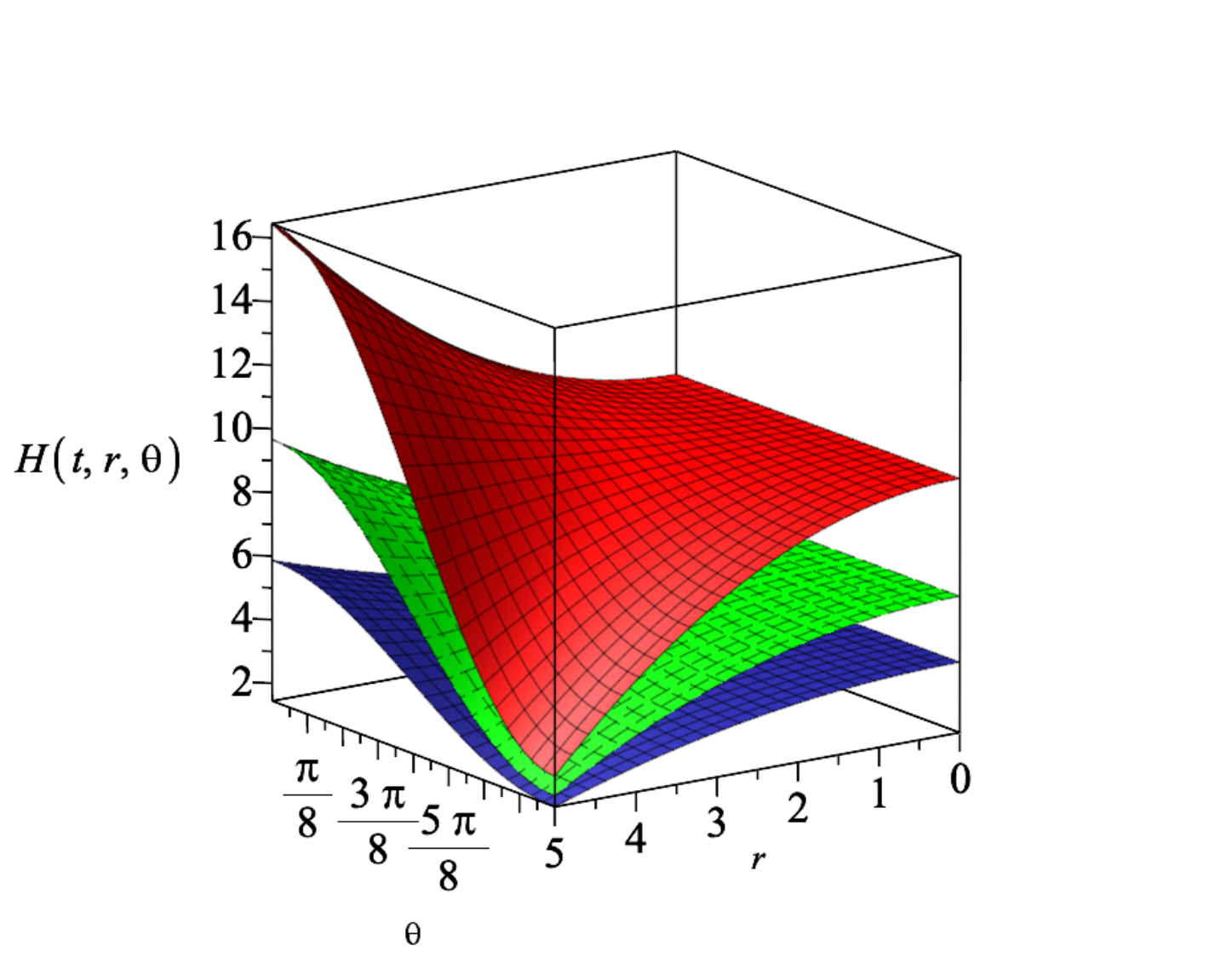}
    \caption{The metric function $H(t,r,\theta)$ in $(4+1)$-dimensions, where the upper surface, middle surface and the lower surface correspond to specific time slices $t=1$, $t=2$ and $t=3$, respectively. We consider the constants as $a=1$, $g_+=0.5$, $g_-=15$ and $\Lambda=1$. }
    \label{figure:Hz}
\end{figure}

In figure \ref{figure:Hz}, we show the behaviour of the metric function $H(t,r,\theta)$ in 5-dimensional spacetime ($N=4$) for three different time slices.

\begin{figure}[h]
	\centering
		\begin{subfigure}{0.4\linewidth}
		\includegraphics[width=\linewidth]{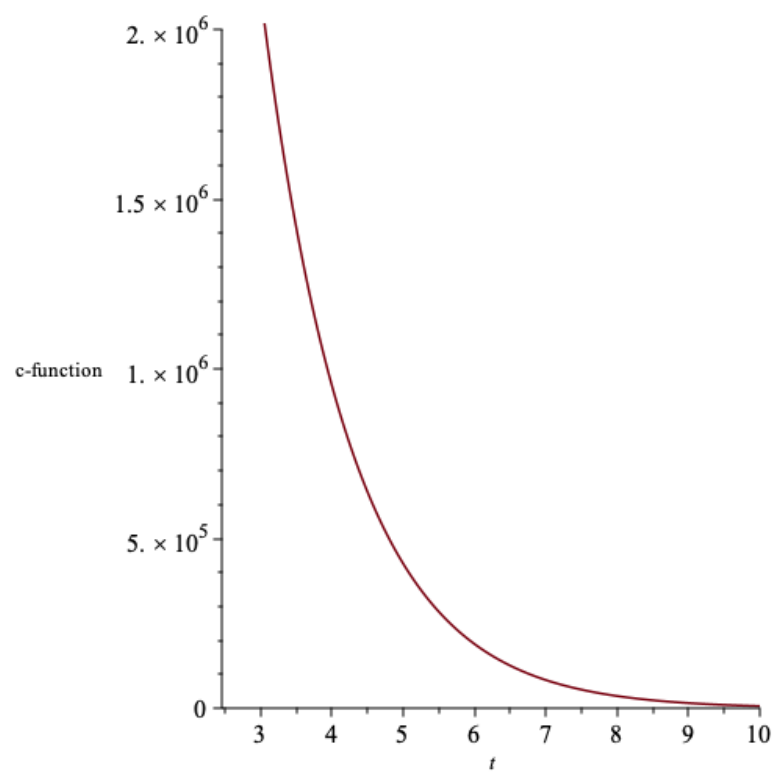}
		\caption{\centering}
		\label{fig:2c}
	\end{subfigure}
	\begin{subfigure}{0.4\linewidth}
		\includegraphics[width=\linewidth]{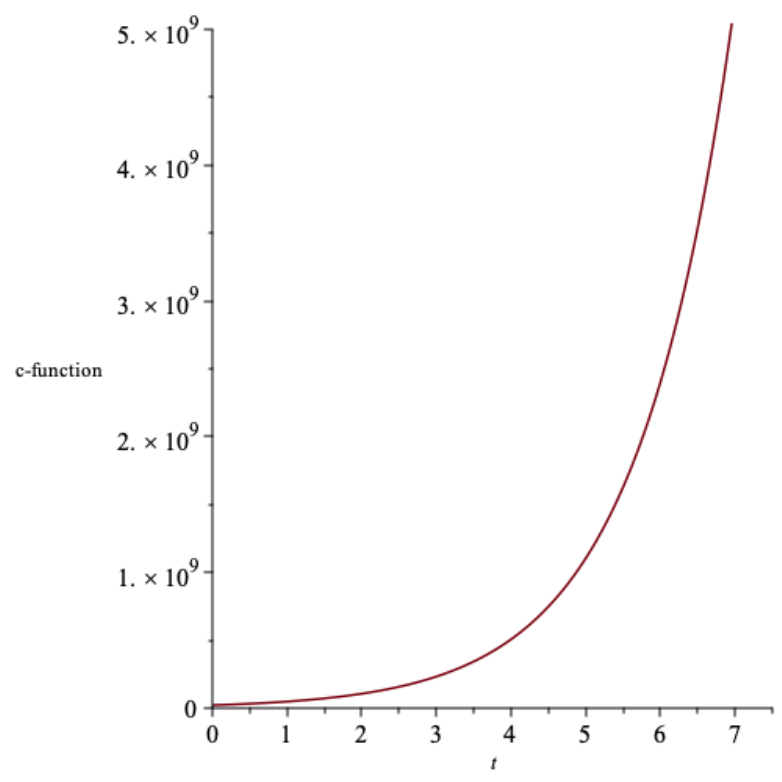}
		\caption{\centering} 
		\label{fig:2d}
	\end{subfigure}
	\caption{ $c$-function in terms of time for (figure a) $\epsilon=+1$ and (figure b) $\epsilon=-1$, for a set of values for the constants.}
	\label{fig:cfunc}
\end{figure}

The $c$-function in $N+1$-dimensions is given by
\begin{equation}
    c\sim \frac{1}{(G_{tt})^{\frac{N-1}{2}}},
\end{equation}
where $G_{tt}$ is the effective Einstein field tensor. For $N=4$, we show the behaviour of the $c$-function in terms of the time coordinate, for both $\epsilon=+1$ and $\epsilon=-1$ in figure \ref{fig:cfunc}.

Furnished with our results, we find the electric fields produced by the electromagnetic gauge field (\ref{Az}) as
\begin{equation}
   E_r= -\frac{rg_+\cos{\theta}\sqrt{2N-2}\exp{(\frac{(N+2)}{\sqrt{N^2-N}}\epsilon\Lambda^{1/2}t)}}{\sqrt{N-2}\Bigl((g_+r^2\cos{\theta}+g_-)\exp{(\frac{2}{\sqrt{N^2-N}}\epsilon\Lambda^{1/2}t)+\exp{(\frac{\sqrt{N}}{\sqrt{N-1}}\epsilon\Lambda^{1/2}t)}} \Bigl)^2},
\end{equation}
\begin{equation}
   E_{\theta}= \frac{r^2g_+\sin{\theta}\sqrt{2N-2}\exp{(\frac{(N+2)}{\sqrt{N^2-N}}\epsilon\Lambda^{1/2}t)}}{2\sqrt{N-2}\Bigl((g_+r^2\cos{\theta}+g_-)\exp{(\frac{2}{\sqrt{N^2-N}}\epsilon\Lambda^{1/2}t)+\exp{(\frac{\sqrt{N}}{\sqrt{N-1}}\epsilon\Lambda^{1/2}t)}} \Bigl)^2}.
\end{equation}

\begin{figure}[h]
	\centering
		\begin{subfigure}{0.49\linewidth}
		\includegraphics[width=\linewidth]{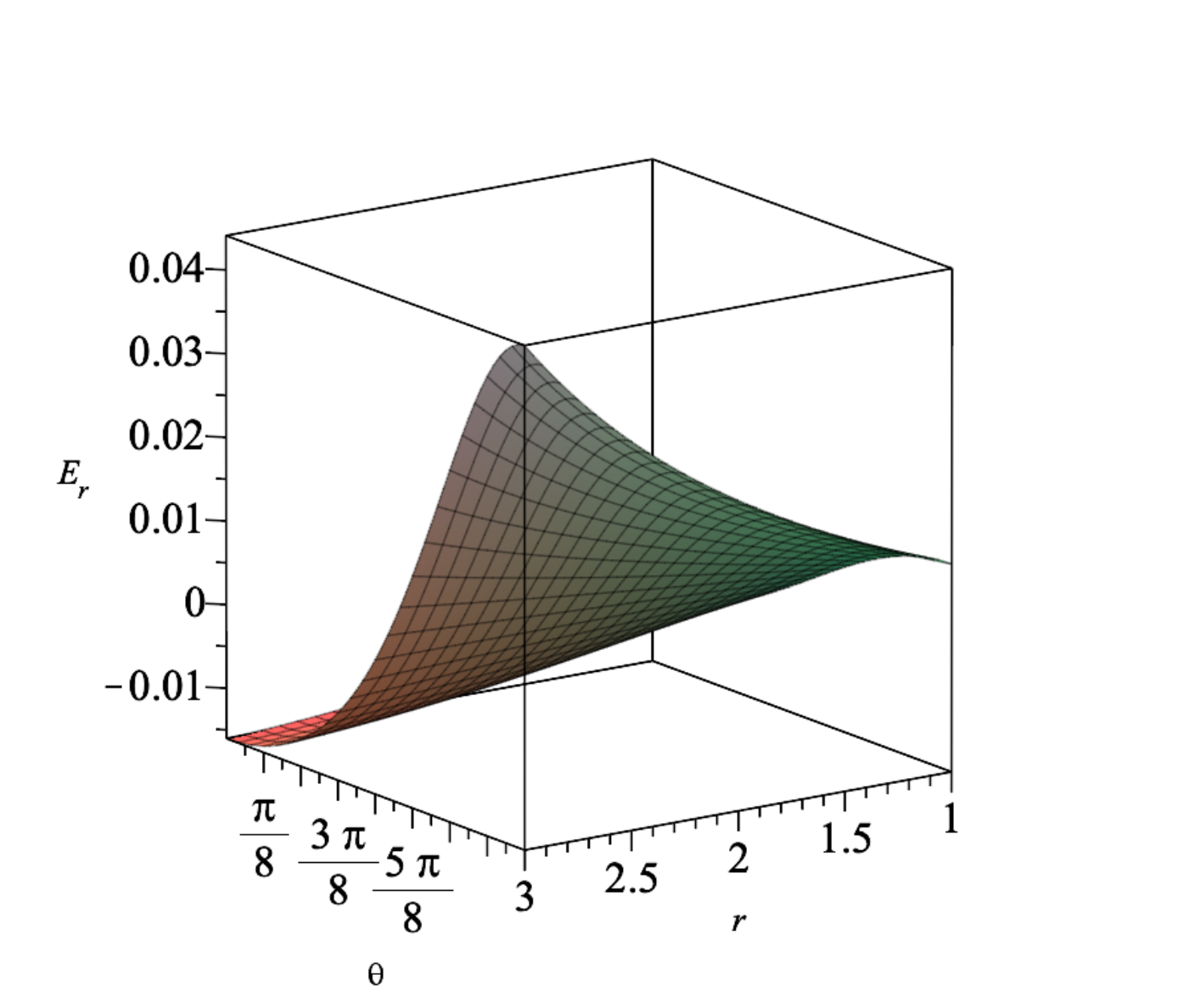}
		\caption{\centering}
	\end{subfigure}
	\begin{subfigure}{0.49\linewidth}
		\includegraphics[width=\linewidth]{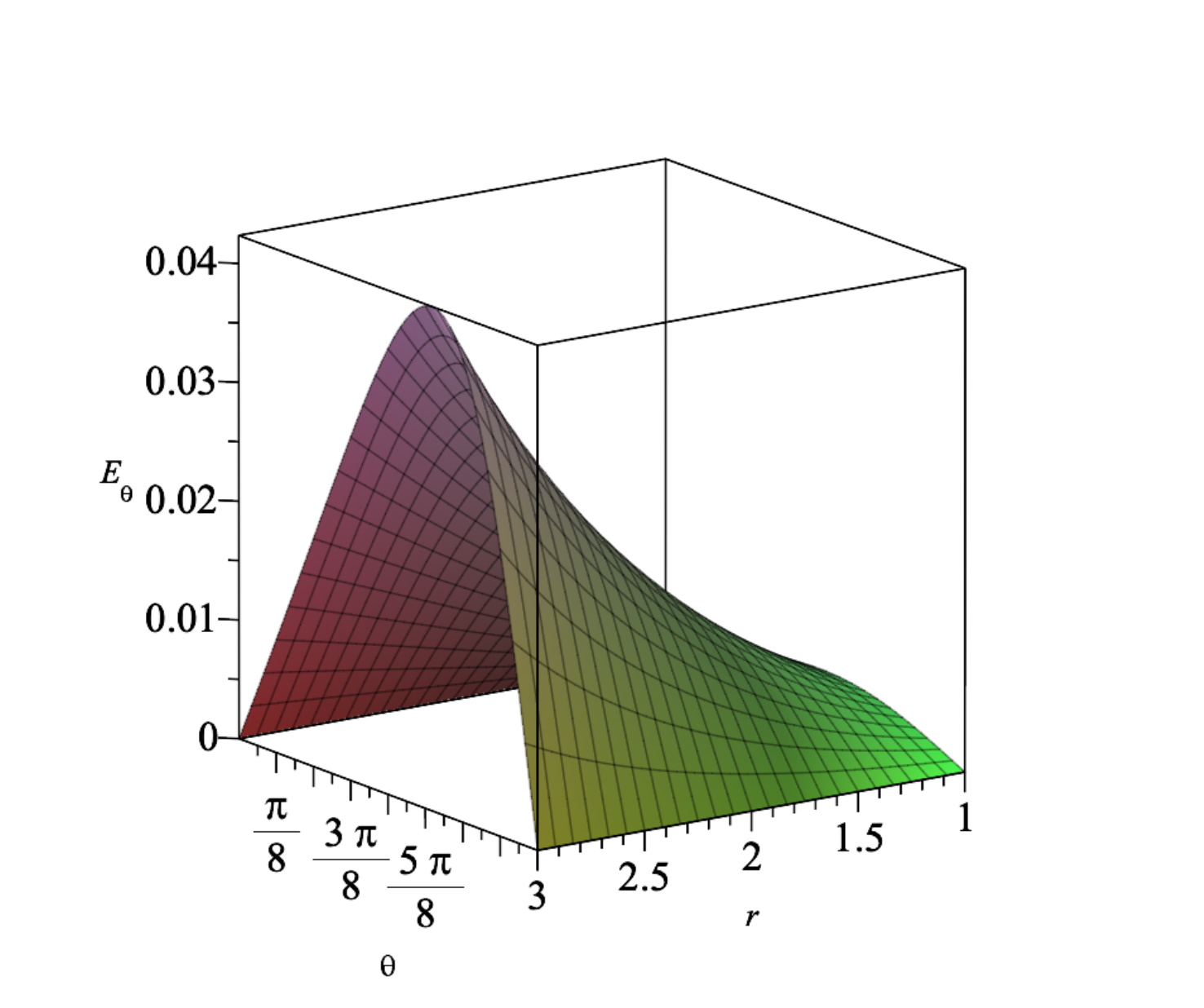}
		\caption{\centering} 
	\end{subfigure}
	\caption{ The electric fields (figure a): $E_r$ and (figure b): $E_{\theta}$ in $(4+1)$-dimensions, where we set $t=2$, $\Lambda=1$, $\epsilon=1$, $g_+=0.5$ and $g_-=15$.}
	\label{fig:Ez}
\end{figure}

In figure \ref{fig:Ez}, we show the behaviour of the electric fields $E_r$ and $E_{\theta}$ in $(4+1)$-dimensions, where we set $t=2$ and $\Lambda=1$.

\section{Embedding of the solutions in higher-dimensional theory}
\label{sec:up}
In this section, we consider the $D$-dimensional gravity coupled to a form field, in presence of a cosmological constant $\Lambda_D$ \cite{ghezelbash2017new},
\begin{equation}
    S_{D}=\int d^Dx\sqrt{-g}(R-\frac{1}{2(q+2)!}\mathcal{F}^2_{[q+2]}+2\Lambda_D), \label{uac}
\end{equation}
where $D=p+q+1$. The $\mathcal{B}_{[q+1]}$ denotes a $q+1$-potential. We note that in equation (\ref{uac}), $R$ is the $D$-dimensional Ricci scalar, and $\mathcal{F}_{[q+2]}$ is given as,
\begin{equation}
    \mathcal{F}_{[q+2]}=d\mathcal{B}_{[q+1]}.\label{FF1}
\end{equation}
Note that in (\ref{FF1}), $\mathcal{F}_{[q+2]}$ is the $q+2$-field strength form, and $d\mathcal{B}_{[q+1]}$ is the exterior derivation of the potential $\mathcal{B}_{[q+1]}$.

We now consider the dimensional reduction of the $D$-dimensional theory (\ref{uac})  to $p+1$-dimensions. We consider an internal curved $q$-dimensional space, with the line element $d\mathcal{K}^2_q$ \cite{ghezelbash2017new}.  We consider the following ansatz for the $D$-dimensional metric,
\begin{equation}
    ds^2_D=e^{-\delta\phi'}ds^2_{p+1}+e^{\phi'(\frac{2}{\delta(p-1)}-\delta)}d\mathcal{K}^2_q, \label{ddim}
\end{equation}
and also the $q+1$-potential $\mathcal{B}_{[q+1]}$,
\begin{equation}
    \mathcal{B}_{[q+1]}=\mathcal{A}_{[1]}\wedge d\mathcal{K}_q. \label{qpo}
\end{equation}

It turns out that the theory in $p+1$ dimensions, is given by \cite{ghezelbash2017new},
\begin{equation}
    S_{p+1}=\int d^{p+1}x(R'-\frac{1}{2}(\nabla\phi')^2-\frac{1}{4}e^{\gamma\phi'}\mathcal{F}^2_{[2]}+2\Lambda_De^{-\delta\phi'}+2\Lambda'e^{-\frac{2}{\delta(p-1)}\phi'}). \label{pdimon}
\end{equation}
We note that in equation (\ref{pdimon}), $R'$ is the $p+1$-dimensional Ricci scalar, and $\Lambda'=R''/2$, where $R''$ is the $q$-dimensional Ricci scalar of the internal space. We also note that in the action (\ref{pdimon}), $\delta$ and $\gamma$ are the dilaton coupling constants, which are given by \cite{gouteraux2011generalized},
\begin{equation}
    \delta=(\frac{2q}{(p-1)(p+q-1)})^{1/2},\label{deltaq}
\end{equation}
\begin{equation}
    \gamma=\delta(2-p).\label{gammap}
\end{equation}
We compare equation (\ref{action}) with  (\ref{pdimon}), and find that the dilaton fields are related by
\begin{equation}
    \phi'=\sqrt{\frac{8}{N-1}}\phi.
\end{equation}
We also find that
\begin{equation}
\Lambda_D=0,
\end{equation}
and
\begin{equation}
2\Lambda '=-\Lambda,
\end{equation}
and
\begin{equation}
    \mathcal{A}_{[1]}=2A_tdt.
\end{equation}

Moreover, the coupling constants in (\ref{pdimon}) are given by
\begin{equation}
    \delta=-\frac{N-1}{2(p-1)b}\sqrt{\frac{8}{N-1}}, \label{deltaa}
\end{equation}
\begin{equation}
    \gamma=-\frac{4a}{N-1}\sqrt{\frac{N-1}{8}},\label{gammaa}
\end{equation}
in terms of coupling constants $a$ and $b$ in (\ref{action}). Equations (\ref{const}), (\ref{gammap}), (\ref{deltaa}) and (\ref{gammaa}) yield
\begin{equation}
p=N.
\end{equation}
Moreover from equation (\ref{deltaq}), we find the dimension of the compact space is given by
\begin{equation}
q=\frac{N-1}{b^2-1}.\label{qq}
\end{equation}
So, we can conclude that our $N+1$-dimensional solutions in section \ref{NOTEQ} with the metric (\ref{st}), the dilaton field (\ref{dil}), and the electromagnetic field (\ref{A}) can be uplifted to the $D=N+1+\frac{N-1}{b^2-1}$-dimensional theory (\ref{uac}) with no cosmological constant $\Lambda_D=0$. Of course, equation (\ref{qq}) imposes another constraint on the coupling constant $b$, to have an integer value for $q$.

\section{Conclusions}
In this article, we present new classes of exact solutions to the $D$-dimensional Einstein-Maxwell-dilaton theory which describes the dynamical black holes in $D\geq 5$ dimensions. The solutions are based partially on the Bianchi type IX geometry, where there are two couplings between the  dilaton field and the electromagnetic field, as well as the dilaton with the cosmological constant. We consider three different cases where the coupling constants are not equal, are non-zero and equal, and finally both are zero. In each case, we use different ansatzes for the $D$-dimensional metric, the dilaton and the Maxwell's field. 
We analytically solve all the field equations and find unique solutions for the metric functions, the dilaton and electromagnetic fields. By solving the field equations, we find that for the case of not equal coupling constants, there is a constraint on the coupling constants. Moreover, we find that the cosmological constant can take any positive, zero or negative values. 
We also show that for special values for the coupling constant $b$, the solutions with two non-equal coupling constants, can be uplifted to a higher dimensional Einstein-form theory with no cosmological constant. Moreover, We  show that for special values for the coupling constant $a$, the solutions with two equal coupling constants, can be uplifted to a higher dimensional Einstein theory with a cosmological constant.
We also should mention that the different classes of exact solutions are completely unique and analytical, and we do not use any approximations to find them, at all. 
We conclude with the observation that the well-known holography between the rotating black holes and the conformal field theories (CFTs) enjoys the independence of the central charges of the CFT on the non-gravitational matter fields \cite{add3, add4}. In this article, we found some exact solutions to the five and higher dimensional gravity coupled to non-gravitational fields. It would be an interesting project to find the rotating versions of the exact solutions, presented in the article. The rotating solutions provide a treasure trove of solutions, including black holes, where we can find and study their holographic dual CFTs. Moreover, we can test the independence of the central charges of the CFT on the non-gravitational fields, for a broader class of gravitational theories.  One other interesting line of research is to seek the possible hidden symmetry in the solutions space of a probe field, in the background of rotating versions of the exact solutions, presented in the article. These symmetries, in general, lead to finding the possible dual hidden CFT to the black holes \cite{add5}. Moreover extending the dual hidden CFT by introducing a deformation parameter in the radial equation of the probe field, as well as finding the different pictures for the dual hidden CFT \cite{add6}, are some of other applications of the rotating versions of the exact solutions, presented in the article.
We leave studying the rotating versions of the exact solutions and their above-mentioned applications in holography for a future article.  

\section{Appendix}
 For the case where the coupling constants are not equal $a\neq b$, we show the $\mathcal{M}_t$ component of the electromagnetic field equation in $N+1$-dimensions, which gives a differential equation for the metric function $H(r,\theta)$

 \begin{eqnarray}
     \mathcal{M}_t &=& -\frac{64(H^{\frac{-N+a^2}{N-2}}\alpha(a^2+N-2)}{(N-2)^2r^5R^4(t)\sin{\theta}\sqrt{16k^4c^4-r^4}\sqrt{16c^4-r^4}}\Bigl(((N-2)\sin{\theta}(c^4k^4-c^4)\cos^2{\psi}
     \nonumber \\
     &-& k^4c^4+r^4/16)Hr^3(\frac{\partial^2}{\partial\theta^2}H)+(N-2)(r^9/32-c^4(k^4+1)r^5/2+8c^8k^4r)\sin{\theta}H
     \nonumber \\
     &\times& (\frac{\partial^2}{\partial r^2}H)+\sin{\theta}a^2(c^4(k^4-1)\cos^2{\psi}-k^4c^4+r^4/16)r^3(\frac{\partial}{\partial\theta}H)^2-(N-2)\cos{\theta}H
     \nonumber \\
     &\times& (c^4(k^4-1)\cos^2{\psi}+c^4-r^4/16)r^3(\frac{\partial}{\partial\theta}H)+4\sin{\theta}(\frac{\partial}{\partial r}H)((a^2r^9/256
     \nonumber \\
     &-& a^2c^4(k^4-1)r^5/16+a^2c^8k^4r)(\frac{\partial}{\partial r}H)-(N-2)(-3r^8/256+c^4(k^4-1)r^4/16
     \nonumber \\
     &+& c^8k^4)H)\Bigl).
 \end{eqnarray}

Solving this equation, we find a solution for $H(r,\theta)$
\begin{equation}
H(r,\theta)=(g_+r^2\cos{\theta}+g_-)^{\frac{N-2}{a^2+N-2}}, \label{H}
\end{equation}
 where $g_{\pm}$ are arbitrary constants.

\vskip 1cm
{\Large Acknowledgments}

This work was supported by the Natural Sciences and Engineering Research
Council of Canada. 

\end{document}